# Title

Extending frequency metrology to increasingly complex molecules: SI-traceable sub-Doppler mid-IR spectroscopy of trioxane

# Authors

Dang-Bao-An Tran,[1,2][†][‡] Mathieu Manceau,[1][†] Olivier Lopez,[1] A. Goncharov,[1][§] Anne Amy-Klein,[1] and Benoît Darquié[1][*]

# Affiliations

1) Laboratoire de Physique des Lasers, Université Sorbonne Paris Nord, CNRS, Villetaneuse, France
2) Faculty of Physics, Ho Chi Minh City University of Education, Ho Chi Minh City, Vietnam
†These authors contributed equally to this work
‡Current address: Time and Frequency Department, National Physical Laboratory, Teddington, United Kingdom
§Permanent address: Institute of Laser Physics, Siberian Branch of the Russian Academy of Sciences, Novosibirsk, Russia

# Abstract

Bringing increasingly complex polyatomic molecules within reach of precision measurement experiments offers fascinating and far-reaching prospects ranging from Earth sciences and astrophysics, to metrology and quantum sciences. Here, we demonstrate sub-Doppler spectroscopic measurements in the mid-IR fingerprint region of, to our knowledge, the largest molecule to date. To this end, we use a high-resolution ~10.3 µm spectrometer based on a sub-Hz quantum cascade laser remotely calibrated against state-of-the-art primary frequency standards *via* a metrology-grade fibre link. We perform saturated absorption spectroscopy in the $\nu_5$ CO stretching mode of 1,3,5-trioxane, $(H_2CO)_3$, at a resolution of ~100 kHz, allowing us to measure the absolute frequency of hundreds of rovibrational transitions at unprecedented uncertainties for such a complex species, as low as ~5 kHz. Our work demonstrates the extension of frequency metrology methodologies to ever larger molecular system, confirming the potential of the technologies we develop for bringing increasingly complex species within reach of ultra-precise measurement experiments.

# Introduction

The mid-infrared (mid-IR) spectral range, spanning 3 to 25 µm and referred to as the molecular fingerprint region, contains narrow and intense rovibrational signatures of a wide variety of molecular species, from the simplest to the most complex. Sub-Doppler high-precision mid-IR rovibrational frequency measurements allowing to reach sub-megahertz resolutions and rovibrational frequency uncertainties at the kilohertz level or better (~$10^{-10}$ or better relative uncertainties) have so far been limited to a handful of relatively simple – typically less than 5 atoms – and symmetric species ($CO_2$ (*1*), OCS (*2*), $O_3$ (*3*), $PH_3$ (*4*), $NH_3$ (*5*), $CH_4$ and methyl-halides (*6*, *7*), $SiF_4$ (*8*), $OsO_4$ (*9*), $C_2H_4$ (*10*), $SF_6$ (*11*)) and a few more "exotic" molecules (HCOOH (*12*) and $CH_3OH$ (*13*, *14*), two asymmetric tops, CHFClBr (*15*), a chiral species, $CH_3ReO_3$ (*16*), a 9 atoms heavy organo-metallic compound). Extending these measurements to increasingly complex molecules has far-reaching applications ranging from Earth sciences and astrophysics, to metrology, fundamental physics measurements and quantum sciences, with a variety of investigations requiring mid-IR frequency uncertainties ranging from $10^{-10}$ to $10^{-15}$. Quantum state resolved spectroscopy at the $10^{-10}$ level of accuracy which surpasses the performance of typical Doppler-limited or Fourier transform infrared (FTIR) spectrometers, is needed for filling molecular databases with increasingly more accurate molecular parameters. Accurate determination of line positions and other molecular parameters (including non-Voigt line shape effects) in laboratory is crucial for atmospheric (*17*, *18*), interstellar or protostellar (*19*) physics and chemistry, in particular for monitoring gas concentrations or searching new molecules by remote sensing experiments. The numerous degrees of freedom and wealth of nearly degenerate hyperfine/rovibrational energy levels available in complex molecular systems also offer unique opportunities for improving tests of fundamental physics. Polyatomic molecules are for instance increasingly being used to test fundamental symmetries, measure spatio-temporal variation of fundamental constants, search for dark matter, … and many of these experiments can be cast as measurements of resonance frequencies of molecular transitions highlighting the importance of frequency metrology. Large amplitude motion such as tunneling inversion or internal rotation in polyatomic molecules enhances for example the sensitivity to variations in fundamental constants (*20*). Mid-IR measurements at the $10^{-12}$ level of uncertainty *e.g.* in methanol would allow constraints on astrophysical variations of the proton-to-electron mass ratio µ to be refined by comparing laboratory data to spectra of cosmic objects (*21*). Further reaching the $10^{-14}$-$10^{-15}$ levels could lead to the tightest direct and model-free constraints thus far on possible current-epoch variations of µ (*11*, *22*, *23*). This is also the level required to probe parity-violating interactions or search for signatures of dark-matter by measuring the tiny energy differences between chiral enantiomers of complex and heavy organo-metallic compounds (*23–26*). Owing to their strong intramolecular interactions, symmetries, and long-lived rotational and vibrational degrees of freedom, large molecules are also promising platforms for probing the many-body dynamics of large spin networks (*27*, *28*) and quantum information processing (*29*, *30*). Quantum state resolved mid-IR spectroscopy and the precise measurement of transition energies between molecular eigenstates is key to unlock these capabilities.

One reason can explain the difficulty to extend frequency metrology to large polyatomic molecules: molecules beyond ~10 atoms typically exhibit reduced signal amplitudes and an increasing degree of spectral congestion in the usual mid-IR bands with increasing atom number as a result of the two following intrinsic effects. First, the increase in both the number of vibrational modes and the magnitude of moments of inertia for every additional atom results in many more rotation-vibration states populated at a given temperature. Second, as the number of vibrational modes increases, bright rovibrational states excited by a mid-IR photon may be strongly coupled to a highly congested manifold of background dark states stemming from lower-lying vibrational modes, leading to intramolecular vibrational redistribution (IVR) (*31*) or other more subtle rovibrational coupling mechanisms (*28*). This represents an obstacle for high-resolution spectroscopy of large molecules as it results in severe spectral congestion, sometimes even obscuring spectra to the point of yielding quasi continuous absorption profiles (*32*). Bringing increasingly large molecules within reach of precision measurements therefore requires us to identify the species that suffer least from intramolecular rovibrational couplings. In this letter, we demonstrate saturation spectroscopic measurements in the $\nu_5$ CO stretching mode of 1,3,5-trioxane, centered at 977 cm$^{-1}$, at a resolution of ~100 kHz. To this end, we use our high-resolution ~10.3 µm spectrometer based on a sub-Hz quantum cascade laser (QCL) remotely calibrated against state-of-the-art primary frequency standards *via* a metrology-grade fibre link (*9*, *13*, *14*). Trioxane, (H$_2$CO)$_3$ shown as an inset in Figure 1, with 12 atoms is to our knowledge the largest molecule for which sub-Doppler mid-IR spectroscopy is demonstrated, and this work achieves an unprecedented level of resolution for such a complex species. Prior to this work, only three Doppler broadened spectroscopic studies of the rotationally resolved vibrational spectrum of trioxane have been reported (in the mid-IR, the $\nu_{16}$ (33) and $\nu_{17}$ (34) bands at 1177 and 1071 cm$^{-1}$, and in the far-IR (35), the $\nu_{19}$, $\nu_7$ and $\nu_{20}$ bands at 525, 466 and 297 cm$^{-1}$). Six rovibrational transitions of the $\nu_5$ CO stretching mode have also been crudely determined with a 40 MHz uncertainty using a $CO_2$ laser (36), and the same report shows a saturated absorption spectrum (the only published to our knowledge) recorded in the vicinity of the 10 μm P(20) $CO_2$ laser emission line, but without any quantitative description or analysis.

Trioxane is a three-unit cyclic polymer of formaldehyde. It is one of the smallest members of a class of molecules called polyoxymethylene (POM, $(H_2CO)_n$). Trioxane is of significant interest across various fields. In astrophysics, with other various forms of POMs, it has been proposed as a source of formaldehyde in cometary comae, and is as such a key molecule in studies of prebiotic chemistry (37, 38). Although it has been studied in laboratory simulations of cometary conditions (37), it has not been unambiguously detected yet in real comets. Additionally, given its rich microwave, THz and mid-IR spectrum, trioxane is an important molecule for the realization of optically-pumped far-infrared gas lasers (36), for which there has recently been renewed interest with the prospects offered by the use of a compact continuously tunable mid-IR QCL as a pumping source instead of a bulky mid-IR gas laser (39, 40). Trioxane also holds promises for mid-IR frequency metrology and calibration given the almost continuous set of transitions that could be used as a grid of references in the 850-1500 cm$^{-1}$ spectral window (41), corresponding to a more than an-order-of-magnitude broader span compared to the currently used $OsO_4$ (42), $SF_6$ (43, 44) and $CO_2$ (1) absolute frequency grids which only covers the 935-1090 cm$^{-1}$ window.

This dense spectrum is also the consequence of the large number of available energy levels that may favour the onset of rovibrational couplings, blur out spectroscopic resolution and hinder the use of this molecule for precision spectroscopic measurements. However, we find spectral line widths to be fully consistent with the expected combination of pressure, transit-time and power broadening, as well as distortion induced by frequency modulation (FM), with no contribution from intramolecular rovibrational coupling at our ~100 kHz resolution. This has allowed us to measure the absolute frequency of about 300 rovibrational transitions over a spectral region of a few 10 GHz, at unprecedented frequency uncertainties for such complex species, as low as ~5 kHz. Amongst a vast amount of lines most likely belonging to hot-bands, a few tens are assigned and used to determine the spectroscopic constants of the first excited vibrational mode with record-low uncertainties. Our work thus demonstrates the extension of frequency metrology methodologies to the largest molecular system to date, confirming the potential of the technologies we develop for bringing increasingly complex species within reach of ultra-precise measurement experiments.

## Results

**Tunable and SI-traceable frequency-comb-stabilised QCL**. Figure 1 gives a schematic overview of our SI-traceable and ultra-stable QCL-based spectrometer. In this work, we only briefly present the key aspects of the setup, and the reader is referred to Refs. (13, 14) for a detailed description (see also details in *Material and Methods* below). The frequency reference operated at LNE-SYRTE is a 1.54 μm fiber laser locked to an ultra-stable cavity whose absolute frequency $\nu_{ref}$ is continuously calibrated to the primary frequency standards of LNE-SYRTE. The resulting ultra-stable laser (USL in Figure 1) exhibits a frequency stability lower than $10^{-15}$ between 0.1 and 10 s and its absolute frequency is known with a relative uncertainty lower than $4\times10^{-14}$ after 1-s averaging time. This optical frequency reference is transferred to LPL *via* a 43-km long fibre link of the REFIMEVE infrastructure (*45*) with active compensation of the propagation-induced phase noise. A 200 MHz ultrastable RF reference $f_{ref}$ derived from a combination of a liquid-helium cooled cryogenic sapphire oscillator and a hydrogen maser, and monitored on the atomic fountains of LNE-SYRTE ($10^{-15}$ stability at 1 s, a few $10^{-14}$ absolute frequency uncertainty (*46*)) is also transferred through the 43-km long link using an intensity modulated auxiliary 1.54 µm laser diode.

At LPL, a ~1.54 µm laser diode is used as an optical local oscillator (OLO, frequency $\nu_{OLO}$). The upper sideband generated by an electro-optic modulator (EOM) is phase-locked to the remote optical reference *via* phase-lock loop $PLL_1$ in Figure 1. Adjusting $f_{EOM}$ (the microwave frequency fed to the EOM) allows the OLO carrier frequency to be continuously tuned over 9 GHz (*13*, *14*). An optical frequency comb (OFC) is then used to transfer the spectral purity, SI-traceability and wide tuneability of the OLO to a 10.3 µm QCL. The comb repetition rate $f_{rep} \sim 250$ MHz is locked to the OLO using phase-lock loop $PLL_2$. Following frequency up-conversion in a non-linear crystal, the QCL frequency is then locked to a high harmonic *n* of $f_{rep}$ using $PLL_3$. The locking chain combining cascaded phase-lock loops allows the spectral purity of the remote reference to be transferred to the QCL, which thus exhibits a frequency stability below $10^{-15}$ for integration time between 0.1 and 10 s and a line width of ~0.1 Hz (*13*, *14*). Scanning the EOM frequency allows $\nu_{OLO}$, $f_{rep}$, and $\nu_{QCL}$ to be consequently tuned in a series of discrete steps. The QCL's absolute frequency is directly traceable to both $\nu_{ref}$ is and $f_{ref}$, respectively the remote 1.54 µm and RF reference. It is determined with a corresponding uncertainty of $4 \times 10^{-14}$ or $10^{-11}$ after 1-s averaging time, depending on the reference used (see details in *Material and Methods* below)

**Broadband saturation spectroscopy**. The SI-traceable sub-Hz QCL is used to carry out saturated absorption spectroscopy of the $\nu_5$ vibrational mode of 1,3,5-trioxane at ~971.5341 cm$^{-1}$ in a multi-pass absorption cell (see details in *Material and Methods* below). Spectra are recorded at room temperature at a typical pressure of 1.5 Pa. Figure 2 (A) (black points) shows a spectrum spanning 255 MHz ranging from 971.527 to 971.536 cm$^{-1}$ exhibiting ten rovibrational resonances. In Figure 2 (B), we zoom in on the $P(16, 4)$ transition (see transition assignments below). Resonances exhibit a peak-to-peak line width of ~400 kHz, limited by a combination of collisional, transit-time and power broadening, as well as FM-induced-distortion. Broadband saturated absorption spectra are obtained by combining several adjacent spectra spanning typically 300 MHz (such as Figure 2 (A)'s spectrum) with a small overlap of a few 10 MHz from one to another. After each individual spectrum, the QCL is unlocked, shifted by ~250 MHz (adjusting the QCL's current), and relocked to the next harmonic of the comb repetition rate (*cf Material and Methods* below). Adjacent spectra are combined resulting in spectra spanning a few gigahertz. Figure 3 (B) shows a sub-Doppler spectrum of the trioxane $\nu_5$ vibrational mode sub-branch $P(J=16, K)$ spanning ~2.7 GHz (black curve, see next section for transition assignments). It is composed of 12 cascaded measurements and exhibits 141 trioxane lines of various strengths, exhibiting signal-to-noise ratios (SNRs) ranging from 10 to 300. We have also recorded spectra of sub-branches $P(J=15, K)$ around 971.9 cm$^{-1}$ and $P(J=17, K)$ around 971.2 cm$^{-1}$, respectively spanning ~1.9 GHz and ~2.7 GHz and exhibiting 69 and 104 transitions. Corresponding spectra are provided in the Supplementary Materials.

**Spectral assignment**. Rovibrational transitions between the ground and the first vibrationally excited state of the $\nu_5$ mode of trioxane are designated by $P(J,K)$, $Q(J,K)$, and $R(J,K)$. This is a parallel mode for which transitions satisfy $\Delta K=0$. The main capital letter $P$, $Q$, and $R$ correspond to $\Delta J= -1$, 0, and 1, respectively, where $J$ and $K$ are the ground state rotational quantum numbers. Transitions are assigned using a model of the vibrational mode resulting from the analysis of: (i) FTIR spectra recorded at the MONARIS Institute (Sorbonne Université, Paris) and at the SOLEIL synchrotron facility; (ii) about 70 saturated absorption transitions in the $P$ and $Q$ branches recorded at the Laboratoire de Physique des Laser (LPL) in a 58-cm long cell using an ultra-stable $CO_2$ laser spectrometer (*47*); and (iii) 47 sub-Doppler $P$ transitions from the present work. This model will be published elsewhere. The resulting simulated stick-spectrum shown in Figure 3 (A) (red lines) for instance allows us to unambiguously assign the $P(16, K)$ $K$-series transitions with $K$ ranging from 0 to 15 among a total of 141 transitions. The numerous unassigned transitions are likely belonging to hot-bands (transitions starting from thermally populated vibrationally excited states of low frequency modes) or to fundamental bands of less abundant isotopologues. They are typically weaker than assigned lines, but some exhibit remarkably high intensities (see red arrows in Figure 3 (B)). For comparison, we also plot in Figure 3 (A) the direct absorption spectrum recorded at a pressure of ~3 Pa using the QCL in free-running regime.

**Spectral line shape and line-center frequency determination**. To achieve a reasonable signal-to-noise ratio, the FM amplitude has been set to 400 kHz, resulting in a distorted line shape compared to a typical Lorentzian profile. Moreover, the residual amplitude modulation associated with FM and the power variation over a scan (resulting from the QCL gain curve, the underlying non-trivial Doppler envelop for such a congested spectrum, multi-pass-cell-induced residual interference fringes not fully averaged out…) can both contribute to an asymmetry of the line shape. We fit our data to a theoretical model for FM spectroscopy inspired by Refs. (*48*, *49*) and already exposed in Ref. (*14*) that takes into account the combined occurrence of intensity and frequency modulation and the associated distortions, after demodulation on any $n^{th}$ harmonic.

To determine the line-center frequency of each transition, we fit to the data our third-harmonic detection line shape model to which we add an offset. Figure 2 (B) shows the resulting fit and corresponding residuals for transition $P(16, 4)$. In our previous works carried out using demodulation on the first harmonic (*13*, *14*), spectra exhibited a baseline that needed to be adjusted with a polynomial. Third-harmonic detection allows us to do away with it (apart from a global offset). This is to be expected, as the most likely baseline contributions (gain curve, Doppler envelop, residual interference fringes) are naturally better filtered out. Details on the data analysis, on the procedure followed for fitting the data and on the uncertainty budget are given in the Supplementary Materials. Table 1 lists the line-centre absolute frequencies and global uncertainties of the 47 rovibrational transitions that we have been able to assign to the three $P(15, K)$, $P(16, K)$, and $P(17, K)$ sub-branches of the $\nu_5$ vibrational mode of trioxane. Line-centre absolute frequencies and uncertainties of the 267 other non-assigned rovibrational transitions of trioxane are also given in the Supplementary Materials. Each transition has been recorded at a single pressure and power. Frequencies and associated uncertainties reported in this work are those determined at this pressure and power (although we cannot determine zero-power and -pressure transition frequencies, we estimate the resulting overall pressure- and power-shift to be smaller than 30 kHz, see details in the Supplementary Materials). We determine central frequencies with uncertainties ranging from 5 kHz to 50 kHz ($10^{-10}$ - $10^{-9}$ in relative value), which is unprecedented for such a large molecule.

**Excited state rotational constants and band center.** We exploit the line-centre frequency measurements to determine the rotational parameters of trioxane in the $v_5 = 1$ vibrationally excited state and the mode centre frequency. Rotational energy terms in the vibrational ground state are given by the following equation:

$$E_{\text{gr}}(J, K) = BJ(J + 1) + (C - B)K^2 - D_J J^2(J + 1)^2 - D_{JK} J(J + 1)K^2 - D_K K^4 + H_J J^3(J + 1)^3 + H_{JK} J^2(J + 1)^3 K^2 + H_{KJ} J(J + 1)K^4 + H_K K^6 \quad (1)$$

with $B$ and $C$ the first order rotational constants; $D_J$, $D_{JK}$, and $D_K$ the quartic centrifugal distortion constants; $H_J$, $H_{JK}$, $H_{KJ}$, and $H_K$ the sextic centrifugal distortion constants. Rotational energy terms in the first vibrationally excited $v_5 = 1$ state are given by:

$$E_{v_5}(J, K) = v_5 + B'J(J + 1) + (C' - B')K^2 - D'_J J^2(J + 1)^2 - D'_{JK} J(J + 1)K^2 - D'_K K^4 + H'_J J^3(J + 1)^3 + H'_{JK} J^2(J + 1)^2 K^2 + H'_{KJ} J(J + 1)K^4 + H'_K K^6. \quad (2)$$

Primed notations are used for rotational constants in the excited state and $v_5$ is the band centre. In the case of $P$ transitions of a parallel band as probed in this work, frequencies are given by:

$$v(J, K) = E_{v_5}(J - 1, K) - E_{\text{gr}}(J, K). \quad (3)$$

To determine rotational constants in the excited state and the band centre frequency, we fix the ground state parameters to those found in the literature (*50*), as listed in Table 2. We then fit the model given in Eq. (7) to the 47 line-center frequencies assigned to the $P(15, K)$, $P(16, K)$, and $P(17, K)$ sub-branches of trioxane listed in Table 1. We recall that the frequencies reported are those measured at 1.5 Pa and at the power used in the measurements (different for the different sub-branches). In order to take into account the corresponding overall pressure- and power-shift, we assign a frequency uncertainty of 30 kHz to all transitions for this fit (see Supplementary Materials). The lines probed in this work are not numerous enough and/or they do not sample appropriately the vibrational band to allow us to fit sextic centrifugal distortion constants which are thus fixed to their ground state values. The range of $J$ values probed (3 only, J = 15, 16 and 17) is even too little to allow a statistically relevant adjustment of the quartic centrifugal distortion constant $D_J$', which is thus also fixed to its ground sate value $D_J$. For parallel bands of top symmetric species, excited state parameters such as $C'$ and $D_K$' cannot be determined independently of their ground state homologues as the corresponding energy terms are $K$-only-dependent (see Eq. (1) and (2)). Only differences $\Delta C = C'-C$ and $\Delta D_K = D_K'-D_K$ have therefore been fitted (see Table 2). The obtained spectroscopic constants of the first excited $v_5 = 1$ vibrational state of trioxane are listed in Table 2. The band center frequency is determined with an uncertainty of $24 \times 10^{-7}$ cm$^{-1}$ (~70 kHz). Our work, compatible with a previous study (*36*), provides improved uncertainties on parameters. It results in particular in a three to four orders-of-magnitude improvement on the determination of the centre frequency and rotational constant $\Delta C$, a factor of 6 improvement on the quartic centrifugal distortion constant $D'_{JK}$ and the determination of a new quartic centrifugal distortion constant, $\Delta D_K$. The standard deviation and root mean square of the differences between experimental and calculated frequencies (listed as Obs.-Calc. in Table S2 of the Supplementary Materials) is about 8 kHz (the Obs.-Calc mean is ~200 Hz). Even if it is ~3 times smaller than the 30 kHz uncertainty assigned to all transitions, the latter accounts for potential pressure- and power-induced systematic shifts and must be used for an appropriate estimation of molecular parameters' uncertainties.

## Discussion

We have demonstrated mid-IR sub-Doppler spectroscopy for, to our knowledge, the largest molecule to date. Measured line shapes that are fully compatible with the expected combination of pressure, transit-time and power broadening, as well as frequency-modulation-induced distortions. Our data thus do not show any signature of intramolecular coupling at play between the probed bright state and the bath of dark rovibrational states that would manifest as additional distortions or the presence of multiplets. This is not necessarily a surprise for a rigid (the cyclic structure of trioxane leads to a relatively rigid nuclear framework, further corroborated by the fact that no higher than sextic centrifugal distortion terms are required for modeling the ground vibrational state (*50*), *cf* Table 2) and symmetric molecule such as trioxane probed at a wavelength of 10 µm corresponding to a relatively low internal energy. However, the present work (and our previous results on methyltrioxorhenium (*16, 24, 47, 51, 52*), an 8 atoms only but much heavier species of interest for probing parity-violating interactions) shows the potential of the extension of frequency metrology to ever larger polyatomic molecules and opens perspectives for exploring the subtlest intramolecular coupling mechanisms, in particular for identifying increasingly large species amenable to precision measurements. Note that a further factor of 10 to 100 gain in resolution is possible by carrying out sub-Doppler spectroscopy in one of our 1.5-m (*9*) or 3-m (*5*) long Fabry-Perot cavity rather than in a multipass cell. Our expectation is that these technologies will enable the exploration of even more complex molecular systems such as polycyclic aromatic hydrocarbons and related species at sub-Doppler resolutions, provided they are combined with the supersonic (*47, 53*) or buffer-gas (*24, 28, 32, 54*) cooling methods (to minimize the number of occupied rovibrational levels and thus enhance signal amplitudes), and probed at potentially even longer wavelengths (*55, 56*).

Our new high-sensitivity, high-resolution and broadband data have furthermore led to the definite assignment of 47 lines amongst the 314 probed. This has allowed us to improve our understanding of the rovibrational structure of trioxane and to build an accurate spectroscopic model of this species. Rotational parameters in the $v = 1$ vibrationally excited state and mode centre frequency have been determined either for the first time or with orders-of-magnitude improved uncertainties compared to the only previous study of trioxane's CO stretching mode (*36*). Precise characterizations and accurate models of such large polyatomic molecules are scarce. They are however essential whether it be for obtaining a good understanding of the level structure for atmospheric quantification, or for selecting the most sensitive transitions/molecules to probe fundamental processes. Here, only a few dozen line positions determined with record 5 to 30 kHz uncertainties have enabled us to build a model supported by molecular parameters with competitive uncertainties compared to what would typically be achieved with ~1000 frequencies known to within ~10 MHz by FTIR spectroscopy. This highlights the importance of extending frequency metrology methods for extracting very accurate molecular parameters of increasingly complex species. The spectral coverage of our QCL has only allowed transitions in the *P* branch to be probed. Using an additional neighbouring QCL would enable us to also sample the *Q* and *R* branches, avoiding potential biases in the determination of molecular parameters. Moreover, probing a few more well-chosen lines spread over the entire vibrational mode, including especially a fraction of high-*J* and -*K* transitions, would certainly lead to the determination of excited state sextic centrifugal distortion constants. Another promising approach for building a model of the highest accuracy consists in complementing a high-resolution FTIR spectroscopic study providing thousands of lines covering the entire vibrational landscape from low to high *K* and *J* with a few ultra-precise sub-Doppler measurements such as those demonstrated in this work, sensibly-chosen to allow the various molecular parameters to be efficiently constrained. We are currently exploring this approach which will be presented elsewhere.

As illustrated in Figure 2 and 3, and S1 and S2 of the Supplementary materials, a fascinating aspect of our experiment lies in the huge amount of new spectroscopic information it provides. In addition to the 47 transitions assigned to the CO stretching mode of trioxane, a total of 267 typically weaker transitions have been resolved and remain unassigned. Those most likely belong to hot-bands (transitions starting from thermally populated vibrationally excited states of low frequency modes, such as (*41*) $\nu_{19}$ at 305 cm$^{-1}$, $\nu_7$ at 467 cm$^{-1}$, or $\nu_{18}$ at 525 cm$^{-1}$) or to fundamental bands of less abundant isotopologues, the study of which is beyond the scope of the present paper. Most of them are much weaker than assigned lines and our data thus also demonstrate the combination of high detection-sensitivity and high resolution that our spectrometer provides. The use of a multipass cell allows such weak hot-bands to be probed at the low pressures required for ultra-high resolution sub-Doppler measurements. Some unassigned lines marked with a red arrow in Figure 3 (B), S1 and S2 exhibit remarkably high intensities. Signatures of the presence of those are in fact clearly visible in the direct absorption Doppler broadened spectrum recorded at a pressure of ~3 Pa using the QCL in free-running regime displayed in Figure 3. The very high resolution provided by saturation spectroscopy is however essential not only for resolving all high-intensity transitions of the $\nu_5$ mode, in particular at low values of *K*, but also for fully unravelling the details of the numerous unassigned lines. Frequency metrology approaches can thus complement traditional methods of rovibrational molecular spectroscopy such as FTIR or Doppler broadened absorption spectroscopy that suffer from lower frequency resolution with a view to contributing to the compilation of molecular databases, essential for retrieving atmospheric concentrations and interpreting astrophysical spectra. The advantage of sub-Doppler spectroscopy is more evident for relatively heavy atmospheric species such as $SF_6$, $ClONO_2$, $CF_4$… which with low-lying vibrational modes exhibit, like trioxane, dense rotational structures and many hot-bands, impossible to resolve by FTIR or Doppler spectroscopy, and for which the molecular databases thus remain largely incomplete (*57*). In this work, extending sub-Doppler spectroscopy to the largest molecules to date gives access to a vastly rich rovibrational spectrum of relatively well isolated lines. As trioxane is expected to exhibit such a continuous set of transitions in most of the 850-1500 cm$^{-1}$ spectral window (*41*), it also holds promises for mid-IR metrology and frequency calibration. Carrying out sub-Doppler spectroscopy in one of our few-meter-long Fabry-Perot resonator (*5*, *9*) will allow us to determine absolute frequencies at the sub-100 Hz level resulting in a metrology-grade absolute frequency grid of considerably increased spectral coverage compared to the current $OsO_4$ (*42*), $SF_6$ (*43*, *44*) and $CO_2$ (*1*) metrology grids limited to the 935-1090 cm$^{-1}$ range.

To conclude, we report the extension of sub-Doppler mid-IR spectroscopy to, as far as we know, the largest molecule to date, namely 1,3,5-trioxane, a 12 atoms molecule. This is achieved by operating our recently developed sub-Hz QCL based multi-pas cell spectrometer. An optical frequency comb and a long-haul optical fibre link are used to transfer to a 10.3 µm the spectral purity of a remote 1.54 µm metrology-grade frequency reference calibrated to the atomic fountains of the French metrology institute. This technology combines record spectral purity, SI-traceability, wide tunability and high sensitivity in new and broad regions of the mid-IR, allowing us to extend frequency metrology methods to increasingly complex species. We demonstrate saturation spectroscopy in the $\nu_5$ CO-stretch vibrational mode of trioxane at an unprecedented ~100 kHz level of resolution for such a large species. Cascaded spectra spanning a few gigahertz provide a remarkably huge amount of previously unknown spectroscopic information. We report line-centre frequencies of more than 300 transitions in the *P* branch, including very weak transitions belonging to hot-bands or fundamental bands of low abundance isotopologues. We demonstrate record global uncertainties from few kilohertz to few 10 kHz, $10^{-10}$ to $10^{-9}$ in relative value, depending on the line intensity, an unrivaled level for such a complex molecule. A few tens of lines only have been unequivocally assigned and exploited to determine the spectroscopic constants of the first excited vibrational mode with record-low uncertainties. Our work demonstrates the potential of our experiment for providing new ultra-precise and complete spectroscopic data of larger and larger species, and therefore for building accurate spectroscopic models of polyatomic molecules and compiling molecular databases with increasingly more accurate parameters, for probing fundamental processes such as intramolecular rovibrational coupling in large molecules and for identifying the most promising species amenable to precision measurements for the most demanding fundamental tests. Ultimately, precision spectroscopy of such targets is the first step toward single quantum state preparation and control of large molecular systems, and opens up a new class of mesoscopic platforms to explore quantum science.

**Materials and Methods**

**Widely tunable SI-traceable frequency-comb-stabilized-QCL-based high-resolution mid-IR spectrometer**. As illustrated in Figure 1, the reference signal, generated at LNE-SYRTE (the French time-frequency metrology institute) by a 1.54 µm ultra-stable laser (USL) is transferred to LPL *via* a 43-km long fiber link. A 1.54 µm laser diode is used as an optical local oscillator (OLO, frequency $\nu_{OLO}$). The upper sideband generated by an electro-optic modulator (EOM) is locked to the reference using phase-lock loop PLL$_1$ (offset frequency $\Delta_1 = \nu_{ref} - \nu_{OLO} - f_{EOM} = 75$ MHz with $f_{EOM}$ the microwave frequency fed to the EOM) ensuring a 9-GHz-wide tunability of the OLO carrier. The beat signal between the OLO carrier and the $p^{th}$ nearest optical frequency comb (OFC) tooth is used, after removal of the comb offset frequency $f_0$, to lock the comb repetition rate $f_{rep}$ ~ 250 MHz to the OLO using PLL$_2$ (offset frequency $\Delta_2 = \nu_{OLO} - p \times f_{rep} = 150$ MHz, with $p \sim 780\,000$). A 1.55 µm comb obtained by sum-frequency generation between a custom 1.85 µm output of the OFC and the quantum cascade laser (QCL, frequency $\nu_{QCL}$) beam in a non-linear crystal of AgGaSe$_2$ is beaten with the original 1.55 µm comb, and the resulting signal (at frequency $\Delta_3 = n \times f_{rep} - \nu_{QCL} = 10.5$ MHz with $n \sim 120\,000$) is processed in PLL$_3$ into a correction applied to the QCL's current. The QCL frequency is then directly linked to the remote frequency reference: $\nu_{QCL} = n/p\,(\nu_{ref} - f_{EOM} - \Delta_1 - \Delta_2) - \Delta_3$. Part of the QCL power is directed to a multi-pass absorption cell to perform saturated absorption spectroscopy. In Figure 1, the panels on the right show the in-loop beat-note signals $\Delta_1$, $\Delta_2$, and $\Delta_3$ of the phase-lock loops. The beat-note signal $\Delta_3$ highlights the frequency modulation (FM, see sub-section *Saturation spectroscopy setup* below) of the QCL at 20 kHz with a FM amplitude of 400 kHz.

**QCL's absolute frequency and its uncertainty**. The QCL's frequency is directly traceable to the remote 1.54 µm reference (see Figure 1's caption and the protocol previously detailed in Refs. (*9*, *13*, *14*)). It is determined with a fractional uncertainty of 4 × $10^{-14}$ after 1-s averaging time limited by the uncertainty on $\nu_{ref}$. We have also developed an alternative protocol which does not require to know $\nu_{ref}$ and for which SI-traceability is provided by the RF reference $f_{ref}$ (*13*). The 4$^{th}$ harmonic of the comb's repetition rate is then detected on a fast photodiode, down-converted by mixing with a signal at 980 MHz from a synthesizer and counted (dead-time-free K+K FXE counter operated in Π-type mode). The synthesizer and counter are both referenced to a 10 MHz signal (stability at 1 s and relative frequency uncertainty better than $10^{-11}$ and $10^{-13}$ respectively (*58*, *59*)) synthesized from the remote LNE-SYRTE 200 MHz reference $f_{ref}$, and we simply retrieve the QCL absolute frequency using $\Delta_3$'s expression given above. This results in a sub-300 Hz uncertainty on the QCL's frequency at 1-s measuring time, limited by the 10 MHz signal stability.

**Saturation spectroscopy setup**. The SI-traceable sub-Hz QCL is used to carry out saturated absorption spectroscopy of the $\nu_5$ vibrational mode of 1,3,5-trioxane at ~971.5341 cm$^{-1}$ in a compact 20-cm long multi-pass absorption cell (182 paths, 36.2-m absorption path length). As illustrated in Figure 1 (see also Refs. (*13*, *14*)), the laser beam is split into a pump and a counter-propagating probe beam both coupled into the multi-pass cell. Sub-Doppler signals are detected on a mercury cadmium telluride (MCT) photodetector. The QCL frequency is modulated at 20 kHz with a FM amplitude of 400 kHz by modulating the 10.5 MHz signal to which beat-note $\Delta_3$ is locked *via* phase-lock loop PLL$_3$ (Figure 1). Demodulation on the third harmonic is carried out using a lock-in amplifier. Interference fringes as typically observed with multipass cells, partially averaged out by vibrating/shaking some optics on the setup, are also naturally filtered out in third harmonic detection. In this work, we are interested in measuring resonance frequencies with the highest accuracy. For this reason, in order to compensate for finite-detection-bandwidth-induced frequency shifts (*60*, *61*), the QCL frequency is swept in both directions with increasing and decreasing frequency, and each spectrum consists in the average of such a pair of up and down scans.

# References

1.  A. Amy-Klein, H. Vigué, C. Chardonnet, Absolute frequency measurement of $^{12}C^{16}O_2$ laser lines with a femtosecond laser comb and new determination of the $^{12}C^{16}O_2$ molecular constants and frequency grid. *J. of Mol. Spec.*, 206–212 (2004).

2.  M. Mürtz, P. Palm, W. Urban, A. G. Maki, More Sub-Doppler Heterodyne Frequency Measurements on OCS between 56 and 63 THz. *Journal of Molecular Spectroscopy* **204**, 281–285 (2000).

3.  R. J. Butcher, B. Saubamea, C. Chardonnet, Carbon dioxide laser saturation spectroscopy and the hyperfine structure of monosubstituted ozone $^{16}O^{16}O^{17}O$ and $^{16}O^{17}O^{16}O$. *Journal of Molecular Spectroscopy* **188**, 142–147 (1998).

4.  R. J. Butcher, C. Chardonnet, C. J. Bordé, Hyperfine lifting of parity degeneracy and the question of inversion in a rigid molecule. *Physical Review Letters* **70**, 2698–2701 (1993).

5.  C. Lemarchand, M. Triki, B. Darquié, C. J. Bordé, C. Chardonnet, C. Daussy, Progress towards an accurate determination of the Boltzmann constant by Doppler spectroscopy. *New J. Phys.* **13**, 073028 (2011).

6.  J. L. Hall, J. A. Magyar, "High resolution saturated absorption studies of methane and some methyl-halides" in *High-Resolution Laser Spectroscopy*, K. Shimoda, Ed. (Springer, Berlin, Heidelberg, 1976), pp. 173–199.

7.  O. Acef et al, Absolute frequency measurements with a set of transportable methane optical frequency standards. *Proceedings of the 1999 Joint Meeting of the European Frequency and Time Forum and the International Frequency Control Symposium*, 742–745 (1999).

8. O. Pfister, F. Guernet, G. Charton, C. Chardonnet, F. Herlemont, J. Legrand, $CO_2$ laser sideband spectroscopy at ultra-high resolution. *J. Opt. Soc. Am. B* **10**, 1521–1525 (1993).

9. B. Argence, B. Chanteau, O. Lopez, D. Nicolodi, M. Abgrall, C. Chardonnet, C. Daussy, B. Darquié, Y. Le Coq, A. Amy-Klein, Quantum cascade laser frequency stabilization at the sub-Hz level. *Nature Photonics* **9**, 456–460 (2015).

10. E. Rusinek, H. Fichoux, M. Khelkhal, F. Herlemont, J. Legrand, A. Fayt, Sub-Doppler Study of the $\nu_7$ Band of $C_2H_4$ with a $CO_2$ Laser Sideband Spectrometer. *Journal of Molecular Spectroscopy* **189**, 64–73 (1998).

11. A. Shelkovnikov, R. J. Butcher, C. Chardonnet, A. Amy-Klein, Stability of the Proton-to-Electron Mass Ratio. *Phys. Rev. Lett.* **100**, 150801 (2008).

12. F. Bielsa, K. Djerroud, A. Goncharov, A. Douillet, T. Valenzuela, C. Daussy, L. Hilico, A. Amy-Klein, HCOOH high resolution spectroscopy in the 9.18 μm region. *J. Mol. Spec.* **247**, 41–46 (2008).

13. R. Santagata, D. B. A. Tran, B. Argence, O. Lopez, S. K. Tokunaga, F. Wiotte, H. Mouhamad, A. Goncharov, M. Abgrall, Y. L. Coq, H. Alvarez-Martinez, R. L. Targat, W. K. Lee, D. Xu, P.-E. Pottie, B. Darquié, A. Amy-Klein, High-precision methanol spectroscopy with a widely tunable SI-traceable frequency-comb-based mid-infrared QCL. *Optica* **6**, 411–423 (2019).

14. D. B. A. Tran, O. Lopez, M. Manceau, A. Goncharov, M. Abgrall, H. Alvarez-Martinez, R. Le Targat, E. Cantin, P.-E. Pottie, A. Amy-Klein, B. Darquié, Near- to mid-IR spectral purity transfer with a tunable frequency comb: Methanol frequency metrology over a 1.4 GHz span. *APL Photonics* **9**, 030801 (2024).

15. T. Marrel, M. Ziskind, C. Daussy, C. Chardonnet, High precision rovibrational and hyperfine analysis of the $\nu_4$=1 level of bromochlorofluoromethane. *J. of Mol. Struct.* **599**, 195–209 (2001).

16. C. Stoeffler, B. Darquie, A. Shelkovnikov, C. Daussy, A. Amy-Klein, C. Chardonnet, L. Guy, J. Crassous, T. R. Huet, P. Soulard, P. Asselin, High resolution spectroscopy of methyltrioxorhenium: towards the observation of parity violation in chiral molecules. *Physical Chemistry Chemical Physics* **13**, 854–863 (2011).

17. O. L. Polyansky, K. Bielska, M. Ghysels, L. Lodi, N. F. Zobov, J. T. Hodges, J. Tennyson, High-Accuracy $CO_2$ Line Intensities Determined from Theory and Experiment. *Physical Review Letters* **114**, 243001 (2015).

18. A. Vaskuri, P. Kärhä, L. Egli, J. Gröbner, E. Ikonen, Uncertainty analysis of total ozone derived from direct solar irradiance spectra in the presence of unknown spectral deviations. *Atmospheric Measurement Techniques* **11**, 3595–3610 (2018).

19. E. Roueff, S. Sahal-Bréchot, M. S. Dimitrijević, N. Moreau, H. Abgrall, The Spectroscopic Atomic and Molecular Databases at the Paris Observatory. *Atoms* **8**, 36 (2020).

20. P. Jansen, H. L. Bethlem, W. Ubachs, Perspective: Tipping the scales: Search for drifting constants from molecular spectra. *The Journal of Chemical Physics* **140**, 010901 (2014).

21. S. Muller, W. Ubachs, K. M. Menten, C. Henkel, N. Kanekar, A study of submillimeter methanol absorption toward PKS 1830−211: Excitation, invariance of the proton-electron mass ratio, and systematics. *A&A* **652**, A5 (2021).

22. G. Barontini, L. Blackburn, V. Boyer, F. Butuc-Mayer, X. Calmet, J. R. Crespo López-Urrutia, E. A. Curtis, B. Darquié, J. Dunningham, N. J. Fitch, E. M. Forgan, K. Georgiou, P. Gill, R. M. Godun, J. Goldwin, V. Guarrera, A. C. Harwood, I. R. Hill, R. J. Hendricks, M. Jeong, M. Y. H. Johnson, M. Keller, L. P. Kozhiparambil Sajith, F. Kuipers, H. S. Margolis, C. Mayo, P. Newman, A. O. Parsons, L. Prokhorov, B. I. Robertson, J. Rodewald, M. S. Safronova, B. E. Sauer, M. Schioppo, N. Sherrill, Y. V. Stadnik, K. Szymaniec, M. R. Tarbutt, R. C. Thompson, A. Tofful, J. Tunesi, A. Vecchio, Y. Wang, S. Worm, Measuring the stability of fundamental constants with a network of clocks. *EPJ Quantum Technol.* **9**, 1–52 (2022).

23. D. M. Segal, V. Lorent, R. Dubessy, B. Darquié, Studying fundamental physics using quantum enabled technologies with trapped molecular ions. *Journal of Modern Optics* **65**, 490–500 (2018).

24. S. K. Tokunaga, R. J. Hendricks, M. R. Tarbutt, B. Darquié, High-resolution mid-infrared spectroscopy of buffer-gas-cooled methyltrioxorhenium molecules. *New J. Phys.* **19**, 053006 (2017).

25. M. R. Fiechter, P. A. B. Haase, N. Saleh, P. Soulard, B. Tremblay, R. W. A. Havenith, R. G. E. Timmermans, P. Schwerdtfeger, J. Crassous, B. Darquié, L. F. Pašteka, A. Borschevsky, Toward Detection of the Molecular Parity Violation in Chiral Ru(acac)$_3$ and Os(acac)$_3$. *J. Phys. Chem. Lett.* **13**, 10011–10017 (2022).

26. K. Gaul, M. G. Kozlov, T. A. Isaev, R. Berger, Chiral Molecules as Sensitive Probes for Direct Detection of P-Odd Cosmic Fields. *Phys. Rev. Lett.* **125**, 123004 (2020).

27. F. Schulz, M. Ijäs, R. Drost, S. K. Hämäläinen, A. Harju, A. P. Seitsonen, P. Liljeroth, Many-body transitions in a single molecule visualized by scanning tunnelling microscopy. *Nature Phys* **11**, 229–234 (2015).

28. L. R. Liu, D. Rosenberg, P. B. Changala, P. J. D. Crowley, D. J. Nesbitt, N. Y. Yao, T. V. Tscherbul, J. Ye, Ergodicity breaking in rapidly rotating C$_{60}$ fullerenes. *Science* **381**, 778–783 (2023).

29. C. M. Tesch, R. de Vivie-Riedle, Quantum Computation with Vibrationally Excited Molecules. *Phys. Rev. Lett.* **89**, 157901 (2002).

30. V. V. Albert, E. Kubischta, M. Lemeshko, L. R. Liu, Quantum theory of molecular orientations: topological classification, complete entanglement, and fault-tolerant encodings. arXiv arXiv:2403.04572 [Preprint] (2024).

31. D. J. Nesbitt, R. W. Field, Vibrational Energy Flow in Highly Excited Molecules: Role of Intramolecular Vibrational Redistribution. *J. Phys. Chem.* **100**, 12735–12756 (1996).

32. P. B. Changala, B. Spaun, D. Patterson, J. M. Doyle, J. Ye, Sensitivity and resolution in frequency comb spectroscopy of buffer gas cooled polyatomic molecules. *Appl. Phys. B* **122**, 292 (2016).

33. B. M. Gibson, N. C. Koeppen, B. J. McCall, Rotationally-resolved spectroscopy of the $\nu_{16}$ band of 1,3,5-trioxane. *Journal of Molecular Spectroscopy* **317**, 47–49 (2015).

34. J. Henninot, H. Bolvin, J. Demaison, B. Lemoine, The Infrared Spectrum of Trioxane in a Supersonic Slit Jet. **68**, 62–68 (1992).

35. C. Richard, P. Asselin, V. Boudon, High-resolution far-infrared synchrotron FTIR spectroscopy and analysis of the $\nu_7$ bands of trioxane. *Journal of Molecular Spectroscopy* **386**, 111614 (2022).

36. D. Dangoisse, J. Wascat, J. M. Colmont, Assignment of laser lines in an optically pumped submillimeter and near millimeter laser: (H$_2$CO)$_3$. *Int J Infrared Milli Waves* **2**, 1177–1191 (1981).

37. H. Cottin, M.-C. Gazeau, J.-F. Doussin, F. Raulin, An experimental study of the photodegradation of polyoxymethylene at 122, 147 and 193 nm. *Journal of Photochemistry and Photobiology A: Chemistry* **135**, 53–64 (2000).

38. H. Cottin, Y. Bénilan, M.-C. Gazeau, F. Raulin, Origin of cometary extended sources from degradation of refractory organics on grains: polyoxymethylene as formaldehyde parent molecule. *Icarus* **167**, 397–416 (2004).

39. A. Pagies, G. Ducournau, J.-F. Lampin, Low-threshold terahertz molecular laser optically pumped by a quantum cascade laser. *APL Photonics* **1**, 031302 (2016).

40. A. Amirzhan, P. Chevalier, J. Rowlette, H. T. Stinson, M. Pushkarsky, T. Day, H. O. Everitt, F. Capasso, A quantum cascade laser-pumped molecular laser tunable over 1 THz. *APL Photonics* **7**, 016107 (2022).

41. M. Kobayashi, R. Iwamoto, H. Tadokoro, Vibrational Spectra of Trioxane and Trioxane-d6. *The Journal of Chemical Physics* **44**, 922–933 (1966).

42. O. Acef, F. Michaud, G. D. Rovera, Accurate determination of OsO$_4$ absolute frequency grid at 28/29 THz. *IEEE Trans. Instr. Meas.* **48**, 567–570 (1999).

43. B. Bobin, C. J. Bordé, J. Bordé, C. Bréant, Vibration-rotation molecular constants for the ground and $\nu_3=1$ states of $^{32}SF_6$ from saturated absorption spectroscopy. *Journal of Molecular Spectroscopy* **121**, 91–127 (1987).

44. O. Acef, C. J. Bordé, A. Clairon, G. Pierre, B. Sartakov, New accurate fit of an extended set of saturation data for the $\nu_3$ band of $SF_6$: comparison of Hamiltonians in the spherical and cubic tensor formalisms. *Journal of Molecular Spectroscopy* **199**, 188–204 (2000).

45. E. Cantin, M. Tønnes, R. L. Targat, A. Amy-Klein, O. Lopez, P.-E. Pottie, An accurate and robust metrological network for coherent optical frequency dissemination. *New J. Phys.* **23**, 053027 (2021).

46. J. Guena, M. Abgrall, D. Rovera, P. Laurent, B. Chupin, M. Lours, G. Santarelli, P. Rosenbusch, M. E. Tobar, R. Li, K. Gibble, A. Clairon, S. Bize, Progress in atomic fountains at LNE-SYRTE. *IEEE Transactions on Ultrasonics, Ferroelectrics, and Frequency Control* **59**, 391–409 (2012).

47. P. Asselin, Y. Berger, T. R. Huet, L. Margulès, R. Motiyenko, R. J. Hendricks, M. R. Tarbutt, S. K. Tokunaga, B. Darquié, Characterising molecules for fundamental physics: an accurate spectroscopic model of methyltrioxorhenium derived from new infrared and millimetre-wave measurements. *Phys. Chem. Chem. Phys.* **19**, 4576–4587 (2017).

48. R. Arndt, Analytical Line Shapes for Lorentzian Signals Broadened by Modulation. *Journal of Applied Physics* **36**, 2522–2524 (1965).

49. S. Schilt, L. Thévenaz, P. Robert, Wavelength modulation spectroscopy: combined frequency and intensity laser modulation. *Appl. Opt., AO* **42**, 6728–6738 (2003).

50. H. Klein, S. P. Belov, G. Winnewisser, Terahertz Spectrum of Trioxane. *Zeitschrift für Naturforschung A* **51**, 123–128 (1996).

51. S. K. Tokunaga, C. Stoeffler, F. Auguste, A. Shelkovnikov, C. Daussy, A. Amy-Klein, C. Chardonnet, B. Darquié, Probing weak force-induced parity violation by high-resolution mid-infrared molecular spectroscopy. *Molecular Physics* **111**, 2363–2373 (2013).

52. B. Darquié, C. Stoeffler, A. Shelkovnikov, C. Daussy, A. Amy-Klein, C. Chardonnet, S. Zrig, L. Guy, J. Crassous, P. Soulard, Progress toward the first observation of parity violation in chiral molecules by high-resolution laser spectroscopy. *Chirality* **22**, 870–884 (2010).

53. B. E. Brumfield, J. T. Stewart, B. J. McCall, Extending the Limits of Rotationally Resolved Absorption Spectroscopy: Pyrene. *J. Phys. Chem. Lett.* **3**, 1985–1988 (2012).

54. B. Spaun, P. B. Changala, D. Patterson, B. J. Bjork, O. H. Heckl, J. M. Doyle, J. Ye, Continuous probing of cold complex molecules with infrared frequency comb spectroscopy. *Nature* **533**, 517–520 (2016).

55. M. Manceau, T. E. Wall, H. Philip, A. N. Baranov, M. R. Tarbutt, R. Teissier, B. Darquié, Demonstration and frequency noise characterization of a 17 μm quantum cascade laser. arXiv arXiv:2310.16460 [Preprint] (2023).

56. Y. Wang, J. Rodewald, O. Lopez, M. Manceau, B. Darquié, B. E. Sauer, M. R. Tarbutt, Wavelength modulation laser spectroscopy of $N_2O$ at 17 μm. arXiv arXiv:2410.22328 [Preprint] (2024).

57. I. E. Gordon, L. S. Rothman, R. J. Hargreaves, R. Hashemi, E. V. Karlovets, F. M. Skinner, E. K. Conway, C. Hill, R. V. Kochanov, Y. Tan, P. Wcisło, A. A. Finenko, K. Nelson, P. F. Bernath, M. Birk, V. Boudon, A. Campargue, K. V. Chance, A. Coustenis, B. J. Drouin, J. –M. Flaud, R. R. Gamache, J. T. Hodges, D. Jacquemart, E. J. Mlawer, A. V. Nikitin, V. I. Perevalov, M. Rotger, J. Tennyson, G. C. Toon, H. Tran, V. G. Tyuterev, E. M. Adkins, A. Baker, A. Barbe, E. Canè, A. G. Császár, A. Dudaryonok, O. Egorov, A. J. Fleisher, H. Fleurbaey, A. Foltynowicz, T. Furtenbacher, J. J. Harrison, J. –M. Hartmann, V. –M. Horneman, X. Huang, T. Karman, J. Karns, S. Kassi, I. Kleiner, V. Kofman, F. Kwabia–Tchana, N. N. Lavrentieva, T. J. Lee, D. A. Long, A. A. Lukashevskaya, O. M. Lyulin, V. Yu. Makhnev, W. Matt, S. T. Massie, M. Melosso, S. N. Mikhailenko, D. Mondelain, H. S. P. Müller, O. V. Naumenko, A. Perrin, O. L. Polyansky, E. Raddaoui, P. L. Raston, Z. D. Reed, M. Rey, C. Richard, R. Tóbiás, I. Sadiek, D. W. Schwenke, E. Starikova, K. Sung, F. Tamassia, S. A. Tashkun, J. Vander Auwera, I. A. Vasilenko, A. A. Vigasin, G. L. Villanueva, B. Vispoel, G.

Wagner, A. Yachmenev, S. N. Yurchenko, The HITRAN2020 molecular spectroscopic database. *Journal of Quantitative Spectroscopy and Radiative Transfer* **277**, 107949 (2022).

58. D. B. A. Tran, "Widely tunable and SI-traceable frequency-comb-stabilised mid-infrared quantum cascade laser : application to high precision spectroscopic measurements of polyatomic molecules," thesis, Université Paris 13, Sorbonne Paris Cité (2019).

59. N. Cahuzac, "Lasers à cascade quantique moyen infrarouges stabilisés sur des peignes de fréquences avec traçabilité au SI : application à la spectroscopie de haute précision du méthanol et de l'ozone," thesis, Université Sorbonne Paris Nord (2023).

60. F. Rohart, S. Mejri, P. L. T. Sow, S. K. Tokunaga, C. Chardonnet, B. Darquié, H. Dinesan, E. Fasci, A. Castrillo, L. Gianfrani, C. Daussy, Absorption-line-shape recovery beyond the detection-bandwidth limit: Application to the precision spectroscopic measurement of the Boltzmann constant. *Phys. Rev. A* **90**, 042506 (2014).

61. F. Rohart, Overcoming the detection bandwidth limit in precision spectroscopy: The analytical apparatus function for a stepped frequency scan. *Journal of Quantitative Spectroscopy and Radiative Transfer* **187**, 490–504 (2017).

**Acknowledgments**

We are grateful to Michel Abgrall, Hector Álvarez Martínez (also affiliated and now at Real Instituto y Observatorio de la Armada, San Fernando, Spain), Rodolphe Le Targat, Etienne Cantin (now at LPL) and Paul-Eric Pottie from Laboratoire Temps Espace (LTE, formerly LNE-SYRTE, Observatoire de Paris - PSL, CNRS, Sorbonne Université, LNE) for developing and operating the 1.54 µm ultrastable frequency reference calibrated to the primary frequency standards of LTE, and providing it through the REFIMEVE optical fiber network. We thank Pierre Asselin from MONARIS (Sorbonne Université, CNRS) for exchanges and discussions in particular in relation to spectroscopic assignments.

**Funding:**

The research leading to these results has received funding from (i) the People Programme (Marie Curie Actions) of the European Union's Seventh Framework Programme (FP7/2007-2013) under REA grant agreement n. PCOFUND-GA-2013-609102, through the PRESTIGE programme coordinated by Campus France; (ii) the Region Île-de-France in the framework of DIM Nano-K and DIM SIRTEQ; (iii) the Agence Nationale de la Recherche projects PVCM (Grant No. ANR-15-CE30-0005-01) and PSYCHe (Grant No. ANR-17-ERC2-0036-01), the LabEx Cluster of Excellence FIRST-TF (ANR-10-LABX-48-01) and the EquipEx Cluster of Excellence REFIMEVE+ (ANR-11-EQPX-0039); (iv) the European Partnership on Metrology, co-financed from the European Union's Horizon Europe Research and Innovation Programme and by the Participating States, through the project (23FUN04 COMOMET); (v) CNRS; (vi) Université Sorbonne Paris Nord. D.B.A. Tran was supported by the Ministry of Education and Training, Vietnam (Program 911).

# Figures and Tables

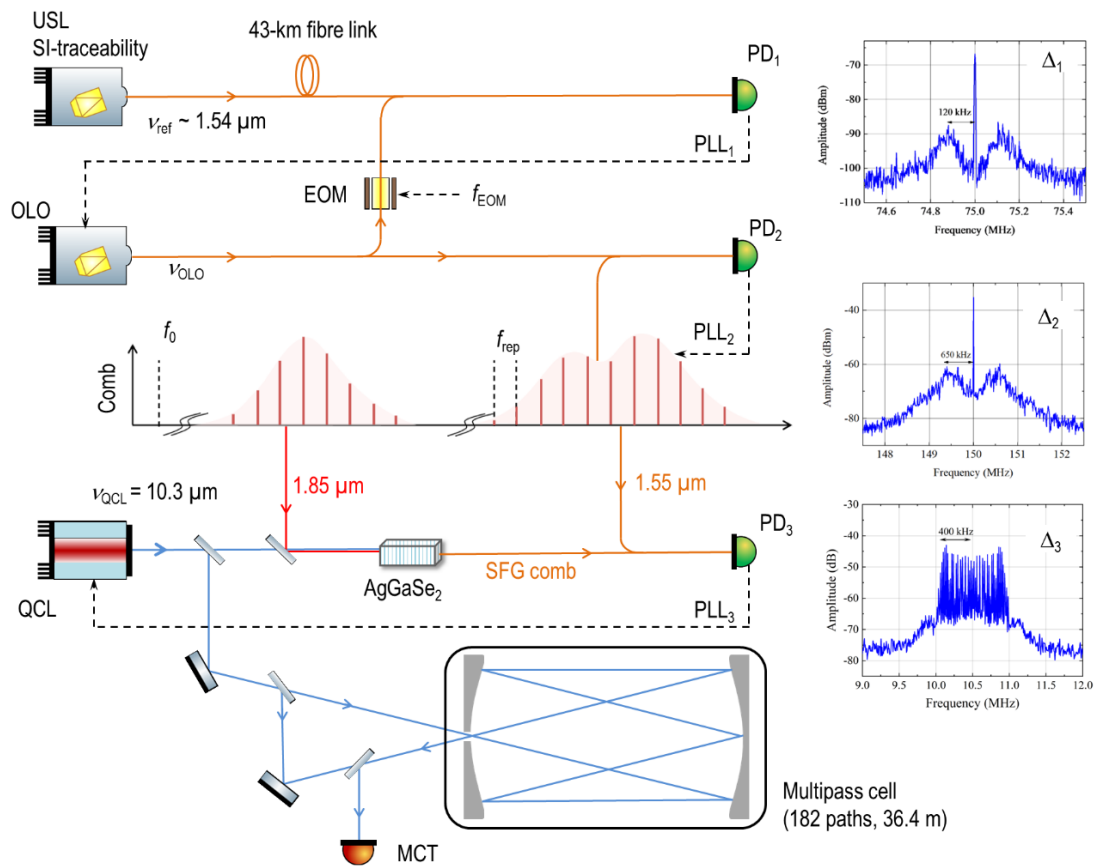

**Fig. 1. Experimental setup**. The reference signal, generated at LNE-SYRTE by a 1.54 µm ultra-stable laser (USL) is transferred to LPL *via* a 43-km long fiber link. A 1.54 µm laser diode is used as an optical local oscillator (OLO, frequency $\nu_{OLO}$). One sideband generated by an electro-optic modulator (EOM) is locked to the reference using phase-lock loop $PLL_1$ ensuring wide tunability of the OLO carrier. The comb repetition rate $f_{rep}$ is locked to the OLO using $PLL_2$. A mid-IR quantum cascade laser (QCL) at 10.3 µm is stabilized to $f_{rep}$ using $PLL_3$. Part of the QCL power is directed to a multi-pass absorption cell to perform saturated absorption spectroscopy. The panels on the right show the in-loop beat-note signals $\Delta_1$, $\Delta_2$, and $\Delta_3$ of the phase-lock loops. $PD_i$: photodetectors, MCT: mercury cadmium telluride photodetector, SFG comb: sum frequency generated comb.

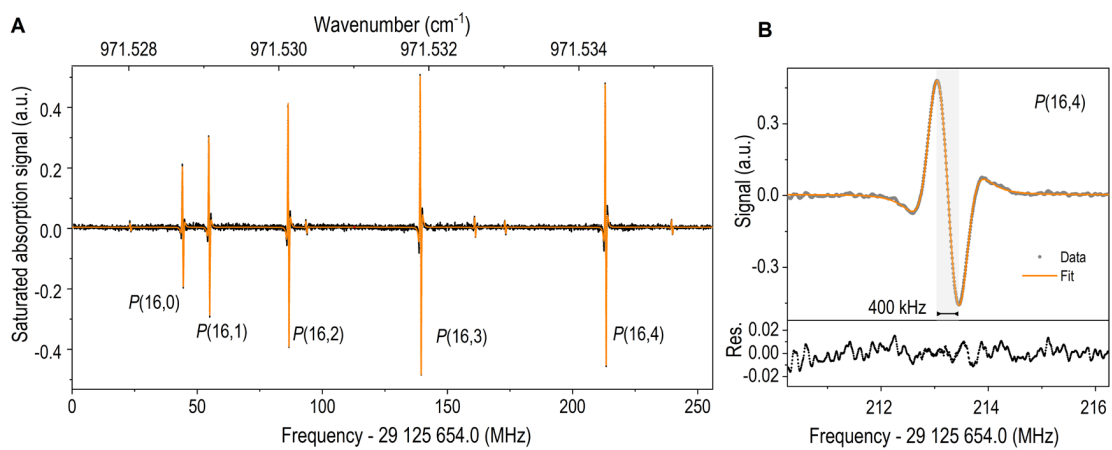

**Fig. 2. Saturated absorption spectrum of trioxane**. (**A**) Saturated absorption spectrum of trioxane (*P*(16, *K*) *K*-series transitions) spanning ~255 MHz recorded using FM and third-harmonic detection, *cf Materials and Methods* (black dots, average of a pair of up and down scans). The orange solid line is a fit to the data corresponding to a sum of ten Lorentzian third derivatives used as a guide-to-the-eye. Experimental conditions: pressure, 1.5 Pa; FM frequency, 20 kHz; FM amplitude, 400 kHz; frequency step, ~7.5 kHz; step duration, 5 ms; lock-in time constant, 10 ms; whole spectrum measurement time, 340 s. (**B**) Zoom showing the *P*(16, 4) line. The orange solid curve is a fit to the data using a model that takes into account residual amplitude modulation associated with FM and the power frequency-dependency over the scan(*14*) (fit reduced chi-square of ~1.1).

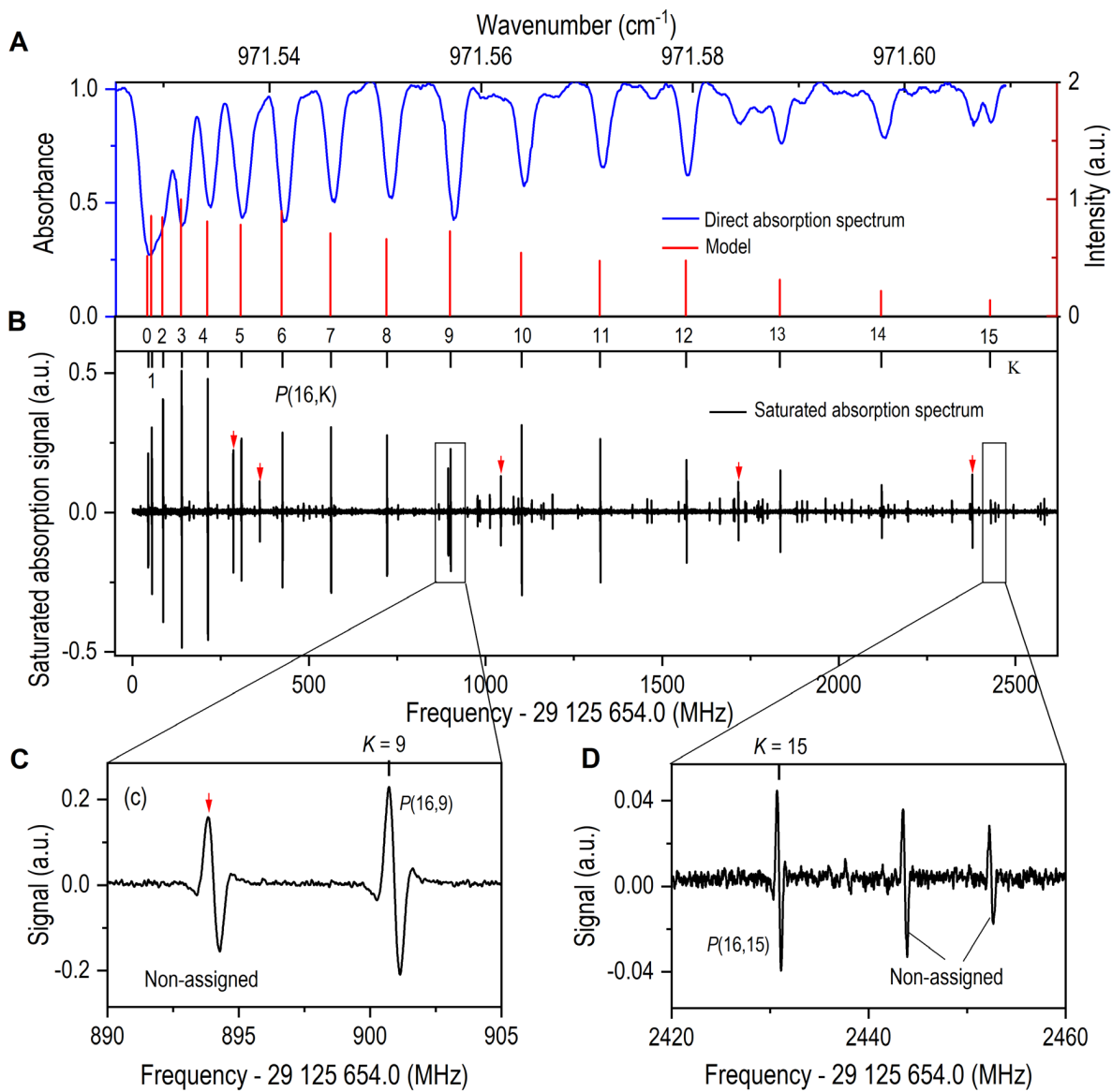

**Fig. 3. Broadband saturation spectroscopy of trioxane**. (**A**) Direct absorption spectrum of the *P*(16, *K*) sub-branch of 1,3,5-trioxane $\nu_5$ vibrational mode recorded at a pressure of 3 Pa together with a stick-spectrum (red sticks) simulated using the model referred to in the text. (**B**) Saturated absorption spectrum spanning ~2.7 GHz. Some relatively high-intensity lines not assigned to the *P*(16, *K*) sub-branch are indicated with red arrows. Experimental conditions: pressure, 1.5 Pa; FM frequency, 20 kHz; FM amplitude, 400 kHz; frequency step, ~7.5 kHz; step duration, 5 ms; lock-in amplifier time constant, 10 ms; average of a pair of up and down scans; whole spectrum measurement time, ~3500 s. Black sticks at the top of the graph indicate the fitted positions of transitions assigned to the *P*(16, *K*) sub-branch and are labelled by their *K* quantum number. (**C**) Zoom on the *P*(16, 9) line lying next to a relatively high intensity unassigned transition. (**D**) Zoom on the *P*(16, 15) line lying next to two unassigned weak transitions.

**Table 1. Assigned trioxane transition frequency positions**. Absolute line-center frequencies and global uncertainties of rovibrational transitions assigned to the *P*(15, *K*), *P*(16, *K*), and *P*(17, *K*) sub-branches of trioxane (at a pressure of 1.5 Pa and at the experimental intra-cell average power, ranging from 0.5 mW to 0.85 mW depending on the region probed, see text and Supplementary Material)

| Transition *P*(15, *K*) | Frequency (kHz) | Transition *P*(16, *K*) | Frequency (kHz) | Transition *P*(17, *K*) | Frequency (kHz) |
|---|---|---|---|---|---|
| *P*(15,0) | 29 136 294 404 (22) | *P*(16,0) | 29 125 698 264.2 (5.6) | *P*(17,0) | 29 115 099 121.9 (6.2) |
| *P*(15,1) | 29 136 304 930 (19) | *P*(16,1) | 29 125 708 818.7 (5.5) | *P*(17,1) | 29 115 109 662.5 (5.8) |
| *P*(15,2) | 29 136 336 683 (19) | *P*(16,2) | 29 125 740 501.3 (5.4) | *P*(17,2) | 29 115 141 271.1 (5.6) |
| *P*(15,3) | 29 136 389 595 (19) | *P*(16,3) | 29 125 793 307.7 (5.4) | *P*(17,3) | 29 115 193 961.1 (5.5) |
| *P*(15,4) | 29 136 463 689 (19) | *P*(16,4) | 29 125 867 252.3 (5.4) | *P*(17,4) | 29 115 267 746.2 (5.5) |
| *P*(15,5) | 29 136 558 994 (19) | *P*(16,5) | 29 125 962 362.4 (5.5) | *P*(17,5) | 29 115 362 658.4 (5.5) |
| *P*(15,6) | 29 136 675 515 (19) | *P*(16,6) | 29 126 078 650.3 (5.5) | *P*(17,6) | 29 115 478 699.6 (5.5) |
| *P*(15,7) | 29 136 813 284 (19) | *P*(16,7) | 29 126 216 153.8 (5.4) | *P*(17,7) | 29 115 615 919.9 (5.5) |
| *P*(15,8) | 29 136 972 346 (19) | *P*(16,8) | 29 126 374 897.0 (5.4) | *P*(17,8) | 29 115 774 324.5 (5.6) |
| *P*(15,9) | 29 137 152 722 (19) | *P*(16,9) | 29 126 554 923.5 (5.4) | *P*(17,9) | 29 115 953 974.5 (6.0) |
| *P*(15,10) | 29 137 354 496 (19) | *P*(16,10) | 29 126 756 284.0 (5.4) | *P*(17,10) | 29 116 154 903.4 (5.6) |
| *P*(15,11) | 29 137 577 673 (20) | *P*(16,11) | 29 126 979 023.2 (5.5) | *P*(17,11) | 29 116 377 183.0 (5.6) |
| *P*(15,12) | 29 137 822 334 (25) | *P*(16,12) | 29 127 223 191.4 (5.5) | *P*(17,12) | 29 116 620 844.5 (5.5) |
| *P*(15,13) | 29 138 088 525 (32) | *P*(16,13) | 29 127 488 856.3 (5.5) | *P*(17,13) | 29 116 885 949.8 (5.5) |
| | | *P*(16,14) | 29 127 776 070.9 (5.7) | *P*(17,14) | 29 117 172 571.1 (5.6) |
| | | *P*(16,15) | 29 128 084 906.9 (6.8) | *P*(17,15) | 29 117 480 752.8 (5.9) |
| | | | | *P*(17,16) | 29 117 810 611.1 (7.5) |

**Table 2. Trioxane spectroscopic parameters**. Spectroscopic constants of the ground state and the first vibrationally excited state of the $\nu_5$ mode 1,3,5-trioxane. The uncertainties in parentheses correspond to one standard deviation.

| Parameters | Ground state $\nu_5 = 0$ | Excited state $\nu_5 = 1$ | |
|---|---|---|---|
| | Klein et al(50) | This work | Dangoisse et al (36) |
| $\nu_5$ (cm$^{-1}$) | - | 977.1709515(24) | 977.172(20) |
| B (MHz) | 5273.257180(33) | 5271.47527(29) | 5271.473(3) |
| ΔC (MHz) | - | 8.97778(98) | 16.2(1.5) |
| $D_J$ (kHz) | 1.3438797(80) | 1.3438797* | 1.352(8) |
| $D_{JK}$ (kHz) | -2.016295(17) | -1.4462(32) | -2.09(2) |
| $ΔD_K$ (kHz) | - | -0.2177(11) | - |
| $H_J$ (mHz) | 0.49061(55) | 0.49061* | - |
| $H_{JK}$ (mHz) | -2.0978(15) | -2.0978* | - |

*Fixed and taken equal to the ground state values.

**Supplementary Text**

**Broadband spectra.** Broadband saturated absorption spectra are obtained by combining several adjacent spectra spanning typically 300 MHz, such as in Figure 3 (B) of the main text that shows a sub-Doppler spectrum of the trioxane $\nu_5$ vibrational mode sub-branch $P(J=16, K)$ spanning ~2.7 GHz and exhibiting 141 trioxane lines. We have also recorded spectra of sub-branches $P(J=15, K)$ around 971.9 cm$^{-1}$ and $P(J=17, K)$ around 971.2 cm$^{-1}$, displayed in Figure S1 and Figure S2, respectively spanning ~1.9 GHz and ~2.7 GHz and exhibiting 69 and 104 transitions.

**Spectral line shape and line-center frequency determination.** To achieve a reasonable signal-to-noise ratio, the FM amplitude has been set to 400 kHz, resulting in a distorted line shape compared to a typical Lorentzian profile. Moreover, the residual amplitude modulation associated with FM and the power variation over a scan (resulting from the QCL gain curve, the underlying non-trivial Doppler envelop for such a congested spectrum, multi-pass-cell-induced residual interference fringes not fully averaged out…) can both contribute to an asymmetry of the line shape. We fit our data to a theoretical model for FM spectroscopy inspired by Refs. (*48,49*) and already exposed in Ref. (*14*) that takes into account the combined occurrence of intensity and frequency modulation and the associated distortions, after demodulation on any $n^{\text{th}}$ harmonic.

To determine the line-center frequency of each transition, we select a ~6 MHz spectral range around each and fit the data to our third-harmonic detection line shape model to which we add an offset. We assign the same experimental error to all data points of a spectrum calculated as the standard deviation of the residuals obtained after fitting a first-order polynomial to a small portion of the spectrum far from resonance. It is roughly the same for all spectra and ranges from 0.28% to 10% of the peak-to-peak amplitude depending on line intensity (corresponding signal-to-noise ratios, SNRs, ranging from 10 to 350). For high-intensity lines (SNR > 40), all line shape parameters are left free in the fit except the FM amplitude fixed to 400 kHz. For lower intensity lines in a given $K$ series only the central frequency, the amplitude and the overall offset are left free, all other parameters being fixed to the average of the corresponding values obtained after fitting high-intensity lines. Figure 2 (B) of the main text shows the resulting fit and corresponding residuals for transition $P(16, 4)$. In our previous works carried out using demodulation on the first harmonic (*13, 14*), spectra exhibited a baseline that needed to be adjusted with a polynomial. Third-harmonic detection allows us to do away with it (apart from a global offset). This is to be expected, as the most likely baseline contributions (gain curve, Doppler envelop, residual interference fringes) are naturally better filtered out.

Table S1 summarizes the uncertainty budget for line centers of trioxane. We have fitted the spectral profile to the data with and without consideration of the asymmetry resulting from residual amplitude variations associated with either FM or the power variation over a scan. There are two independent asymmetry parameters in our model ($B_1$ and $B_2 \times \cos\Psi$ in Eq. (B8) of Ref. (*14*)), respectively related to the quasi-static intensity variations over a scan and to the FM-induced intensity modulation. For each high-intensity transitions (SNR > 40, see above), we have fitted five different models to the data: the complete model with both asymmetry parameters left free; the symmetric model with both asymmetry parameters set to 0; the two models having one asymmetry parameter left free and the other set to 0; a model with the two asymmetry parameters adjusted but constrained to be equal. We have found all fits to converge to the same center-frequency within 17 kHz (root-mean-square dispersion of the differences to the complete model) for lines in the *P*(15, *K*) sub-branch spectral region, and within 1.4 kHz for lines in the *P*(16, *K*) and *P*(17, *K*) sub-branches spectral windows, which we thus assign as a conservative systematic uncertainty resulting from the inaccuracy of our model. These are in fact conservative uncertainties which include the possibility that the asymmetry of the model may or may not be present. The *P*(15, *K*) sub-branch ten times larger uncertainty is due to the more pronounced asymmetry observed in the corresponding data. All spectra have been measured at a pressure of 1.5 ± 0.1 Pa and at an intra-cell averaged power of 0.56 mW, 0.72 mW and 0.84 mW (uncertainty/variation of ~50 µW at most) for *P*(15, *K*), *P*(16, *K*) and *P*(17, *K*) lines respectively (sum of pump and probe power averaged over the 182 paths (*13, 14*)). Our total pressure and power measurement uncertainties result in shift-induced systematic uncertainties <500 Hz and <1 kHz respectively (*13, 14*). The reader is referred to Refs. (*13, 14, 58*) and to the comments given in Table S1 for other details on the systematic effects and the associated line positions uncertainties. Overall, this results in a total systematic uncertainty of 18 kHz for lines recorded in the *P*(15, *K*) sub-branch region (Figure S1), and of 5.4 kHz for lines recorded in the *P*(16, *K*) (Figure 3 of the main text) and *P*(17, *K*) (Figure S2) sub-branch regions.

For most of the 314 transitions studied in this work, a single spectrum was recorded (only 59 transitions have been recorded twice). To determine statistical uncertainties, we simulate experimental spectra spanning 6 MHz. To do this, we employ our line shape model to which we add some experimentally measured noise that consists in the residuals obtained after fitting one of the measured spectra. We numerically simulate 70 such spectra at a given SNR using 70 different experimental residuals. The corresponding statistical uncertainty is estimated by calculating the weighted standard deviation of the 70 fitted center frequencies, with the weights determined from the fits' error bars. As expected, our simulations indicate that the statistical uncertainty in the fitted line center is inversely proportional to SNR and equals 166 kHz for a SNR of 1, in good agreement with the standard deviation of line positions that have been measured twice. We note that carrying out a similar simulation in which we replace the experimentally measured noise by some Gaussian noise with the same variance would result in a 3 times smaller statistical uncertainty, highlighting the non-Gaussian nature of our experimental noise.

Figure S3 displays the systematic, statistical and total frequency uncertainties versus spectrum signal-to-noise ratio. Table 1 of the main text lists the line-centre absolute frequencies and global uncertainties of the 47 rovibrational transitions that we have been able to assign to the three *P*(15, *K*), *P*(16, *K*), and *P*(17, *K*) sub-branches of the $\nu_5$ vibrational mode of trioxane. Line-centre absolute frequencies and uncertainties of the 47 assigned lines and the 267 other non-assigned rovibrational transitions of trioxane are given in Table S2. We also list in Table S2 the amplitude of the corresponding saturated absorption spectroscopic signal in arbitrary units (taken as the signal peak-to-peak fitted amplitude normalized to that of transition *P*(16,3), the most intense fitted line, see Figure 2 (A) and 3 (B) of the main text) and the difference between the observed transition frequencies and those calculated using the spectroscopic constants determined in the main text (see Table 2 in the main text). For the 59 transitions recorded twice, the weighted mean and weighted standard error of the 2 measurements are given. We stress here again that we have conducted measurements at a single pressure (1.5 Pa) and intra-cell average power (ranging from 0.5 mW to 0.85 mW depending on the region probed, see above), and that frequencies and associated uncertainties in Table 1 of the main text and Table S2 are those determined at this pressure and power. We cannot deduce zero-power and -pressure transition frequencies, but we estimate the resulting overall pressure- and power-shift to be smaller than 30 kHz (*13, 14*).

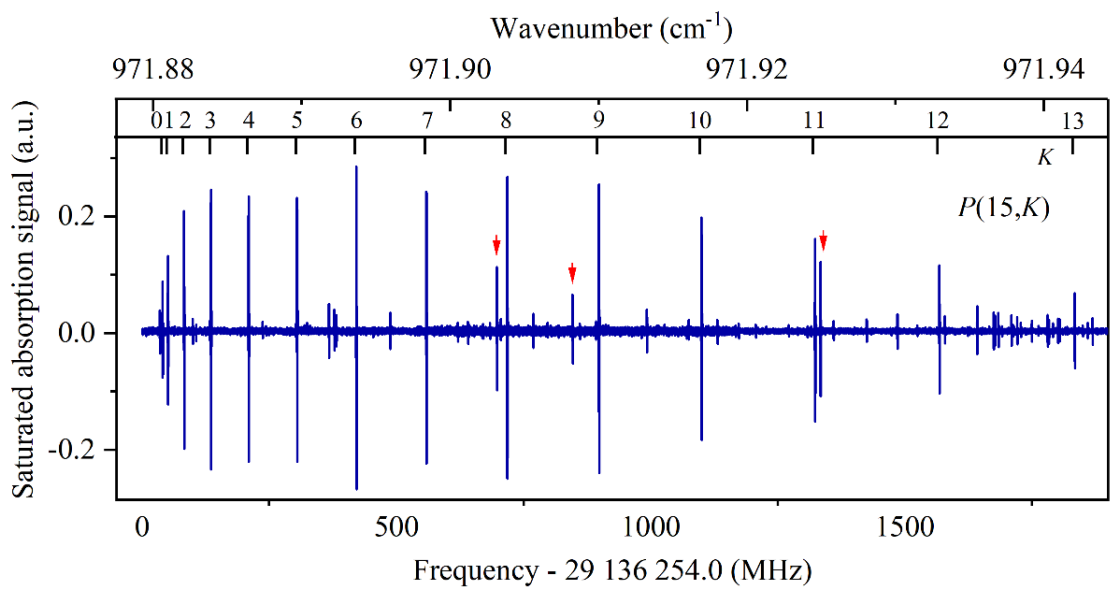

**Fig. S1.**
Saturated absorption spectrum of the P(15, K) sub-branch of 1,3,5-trioxane ν5 vibrational mode spanning ~1.9 GHz. Some relatively high-intensity lines not assigned to the P(15, K) sub-branch are indicated with red arrows. Experimental conditions: pressure, 1.5 Pa; FM frequency, 20 kHz; FM amplitude, 400 kHz; frequency step, ~7.5 kHz; step duration, 5 ms; lock-in amplifier time constant, 10 ms; average of a pair of up and down scans; whole spectrum measurement time, ~2560 s. Black sticks at the top of the graph indicate the fitted positions of transitions assigned to the P(15, K) sub-branch and are labelled by their K quantum number.

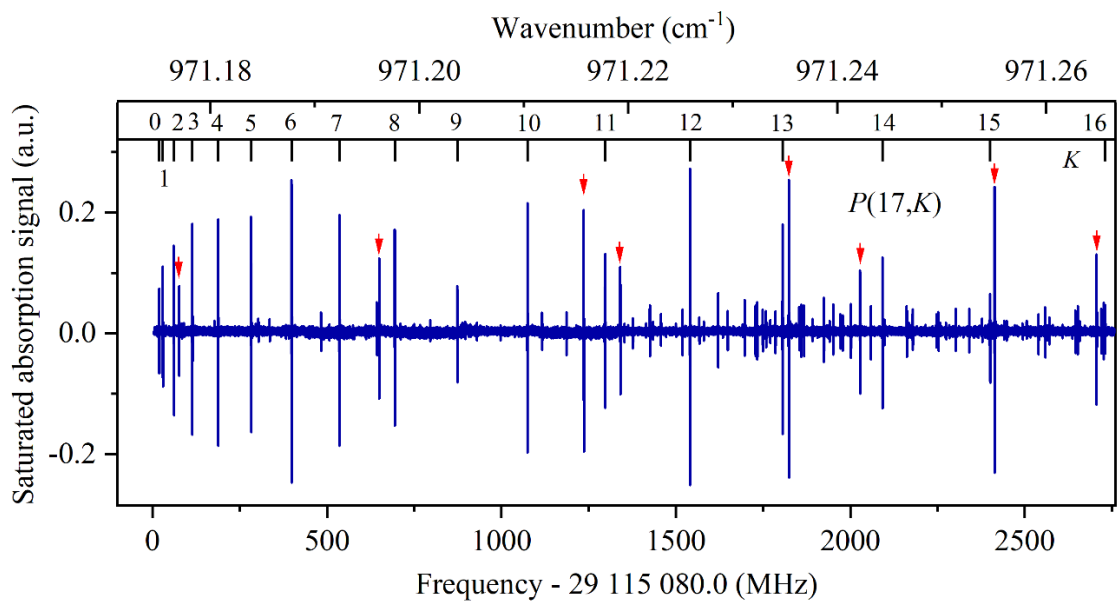

**Fig. S2.**
Saturated absorption spectrum of the P(17, K) sub-branch of 1,3,5-trioxane $\nu_5$ vibrational mode spanning ~2.7 GHz. Some relatively high-intensity lines not assigned to the P(17, K) sub-branch are indicated with red arrows. Experimental conditions: pressure, 1.5 Pa; FM frequency, 20 kHz; FM amplitude, 400 kHz; frequency step, ~7.5 kHz; step duration, 5 ms; lock-in amplifier time constant, 10 ms; average of a pair of up and down scans; whole spectrum measurement time, ~3680 s. Black sticks at the top of the graph indicate the fitted positions of transitions assigned to the P(17, K) sub-branch and are labelled by their K quantum number.

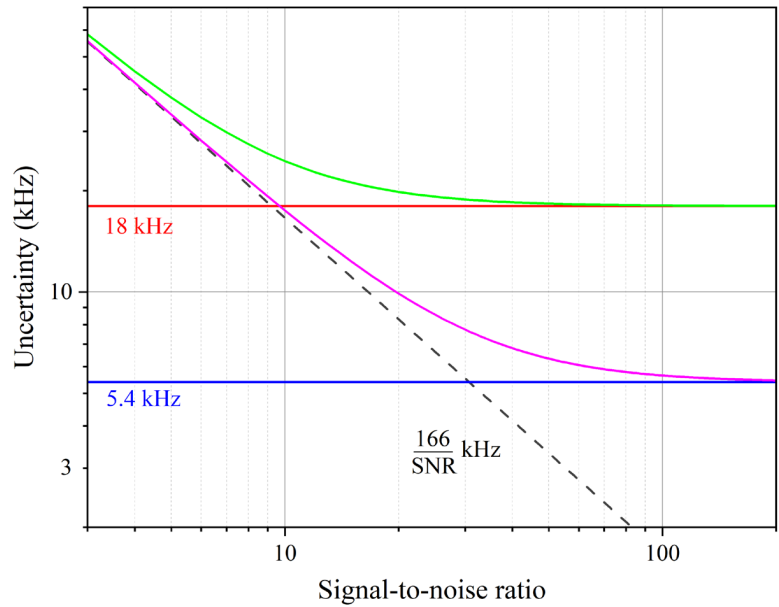

**Fig. S3.**
Frequency uncertainties. Grey dashed line: statistical uncertainty on the fitted central frequency as a function of spectrum signal-to-noise ratio (SNR). Horizontal red and blue lines: systematic-effects-induced uncertainty for lines in the P(15, K) (red), or P(16, K) and P(17, K) (blue) regions. Green and magenta solid lines: total uncertainty versus signal-to-noise ratio for lines in the P(15, K) (green), or P(16, K) and P(17, K) (magenta) regions.

**Table S1.**

Uncertainty budget table for trioxane line position absolute frequencies (*13, 14, 58*).

| Systematic effects | Uncertainty | Comments |
| --- | --- | --- |
| Frequency calibration[a] | ~1 Hz | when traceability to the LNE-SYRTE optical frequency reference $\nu_{\text{ref}}$ is exploited, dominated by the $4 \times 10^{-14}$ uncertainty on $\nu_{\text{ref}}$ |
| | <300 Hz | when traceability to the LNE-SYRTE RF frequency reference $f_{\text{ref}}$ is exploited, limited by a combination of $f_{\text{ref}}$ stability and counter resolution |
| | | (see section *QCL's absolute frequency and its uncertainty* in the main text) |
| Fit model inaccuracy | 17 kHz | for lines in the *P*(15, *K*) sub-branch region |
| | 1.4 kHz | for lines in the *P*(16, *K*) and *P*(17, *K*) regions |
| | | (see text) |
| Pressure / power uncertainties | < 1.2 kHz | pressure and power measurement uncertainties result in shift-induced systematic uncertainties |
| Other spectroscopic effects | 5 kHz | conservative upper bound on the shift induced by gas lens effects, wave-front-curvature-induced residual Doppler shifts, the 2$^{\text{nd}}$-order Doppler shift, Zeeman effects and black-body radiation shift *(13, 14, 58)* |
| Total systematic-effects-induced uncertainty | 18 kHz | for lines in the *P*(15, *K*) sub-branch region |
| | 5.4 kHz | for lines in the *P*(16, *K*) and *P*(17, *K*) regions |
| Statistical uncertainty | 166 kHz/SNR | with SNR the signal-to-noise ratio of the transition |
| Pressure / power shift | <30 kHz | upper limit on the overall power and pressure shift given the experimental conditions at which measurements were conducted |

[a]For six lines in of the *P*(15, *K*) sub-branch region displayed in Figure S1 (including *P*(15, 11) in Table 1 of the main text and other lines marked with an asterisk in Table S2), using traceability to $f_{\text{ref}}$ revealed unexpected fluctuations of $f_{\text{rep}}$, which has led us to assign a conservative 2.5 kHz frequency calibration uncertainty, resulting in the same 18 kHz total systematic-effects-induced uncertainty.

**Table S2.**

Absolute line-center frequencies and global uncertainties of all observed rovibrational transitions of trioxane (at a pressure of 1.5 Pa and at the experimental intra-cell average power, ranging from 0.5 mW to 0.85 mW depending on the region probed, see text). When known, assignments are indicated. Amp.: amplitudes of the corresponding saturated absorption spectroscopic signal (normalized peak-to-peak fitted amplitude, see text) in arbitrary units (a.u.). Obs.-calc.: difference between the observed and calculated (see main text Table 2) transition frequencies.

| P(15, K) sub-branch spectral region | | | | P(16, K) sub-branch spectral region | | | | P(17, K) sub-branch spectral region | | | |
|---|---|---|---|---|---|---|---|---|---|---|---|
| Assignment $P(15,K)$ | Frequency (kHz) | Amp. (a.u.) | Obs.-Calc. (kHz) | Assignment $P(16,K)$ | Frequency (kHz) | Amp. (a.u.) | Obs.-Calc. (kHz) | Assignment $P(17,K)$ | Frequency (kHz) | Amp. (a.u.) | Obs.-Calc. (kHz) |
| | 29136289050 (32) | 0.07 | | | 29125677371.8 (23.9) | 0.03 | | $P(17,0)$ | 29115099121.9 (6.2) | 0.14 | 4.7 |
| | 29136293712 (46) | 0.03 | | $P(16,0)$ | 29125698264.2 (5.6) | 0.40 | -0.8 | $P(17,1)$ | 29115109662.5 (5.8) | 0.20 | 8.8 |
| $P(15,0)$ | 29136294404 (22) | 0.15 | 39.6 | $P(16,1)$ | 29125708818.7 (5.5) | 0.60 | -5.0 | $P(17,2)$ | 29115141271.1 (5.6) | 0.27 | 5.3 |
| $P(15,1)$ | 29136304930 (19) | 0.25 | -13.7 | $P(16,2)$ | 29125740501.3 (5.4) | 0.82 | -1.2 | | 29115156218.3 (6.1) | 0.15 | |
| $P(15,2)$ | 29136336683 (19) | 0.41 | -2.5 | | 29125747660 (16) | 0.04 | | | 29115164433 (23) | 0.02 | |
| | 29136353776 (40) | 0.04 | | $P(16,3)$ | 29125793307.7 (5.4) | 1.00 | -1.4 | | 29115170294 (28) | 0.02 | |
| | 29136360509 (47) | 0.02 | | | 29125815061.5 (8.8) | 0.06 | | $P(17,3)$ | 29115193961.1 (5.5) | 0.35 | -0.2 |
| | 29136387485 (44) | 0.03 | | | 29125827160 (12) | 0.04 | | | 29115258117 (23) | 0.02 | |
| $P(15,3)$ | 29136389595 (19) | 0.48 | -3.0 | $P(16,4)$ | 29125867252.3 (5.4) | 0.80 | -4.4 | $P(17,4)$ | 29115267746.2 (5.5) | 0.37 | -6.9 |
| $P(15,4)$ | 29136463689 (19) | 0.45 | -4.1 | | 29125893776 (15) | 0.05 | | $P(17,5)$ | 29115362658.4 (5.5) | 0.42 | -1.2 |
| | 29136491151 (46) | 0.03 | | | 29125925440 (12) | 0.05 | | | 29115383578 (11) | 0.04 | |
| $P(15,5)$ | 29136558994 (19) | 0.49 | 4.3 | | 29125939359.7 (5.5) | 0.45 | | | 29115415857 (12) | 0.04 | |
| | 29136621733 (33) | 0.09 | | $P(16,5)$ | 29125962362.4 (5.5) | 0.51 | -0.9 | | 29115462370 (24) | 0.02 | |
| | 29136632332 (36) | 0.07 | | | 29125964324 (16) | 0.03 | | $P(17,6)$ | 29115478699.6 (5.5) | 0.44 | -4.6 |
| | 29136636149 (41) | 0.05 | | | 29125999534 (15) | 0.04 | | | 29115563303.4 (9.3) | 0.05 | |
| $P(15,6)$ | 29136675515 (19) | 0.55 | 3.9 | | 29126014211.1 (5.8) | 0.22 | | | 29115586817 (25) | 0.02 | |
| | 29136742200 (39) | 0.06 | | | 29126053315 (19) | 0.03 | | $P(17,7)$ | 29115615919.9 (5.5) | 0.37 | 4.5 |
| $P(15,7)$ | 29136813284 (19) | 0.51 | -1.8 | | 29126060501 (20) | 0.03 | | | 29115654532 (31) | 0.01 | |
| | 29136874963 (51) | 0.03 | | | 29126063484 (16) | 0.03 | | | 29115691394 (22) | 0.02 | |
| | 29136895030 (50) | 0.03 | | $P(16,6)$ | 29126078650.3 (5.5) | 0.56 | -2.2 | | 29115723042.4 (7.5) | 0.07 | |
| | 29136952110 (27) | 0.22 | | | 29126091065 (19) | 0.03 | | | 29115729976.1 (5.7) | 0.21 | |

| | | | | | | | | | | |
|---|---|---|---|---|---|---|---|---|---|---|
| | 29136959455 (50) | 0.03 | | | 29126102960 (29) | 0.02 | | P(17,8) | 29115774324.5 (5.6) | 0.24 | -2.7 |
| P(15,8) | 29136972346 (19) | 0.52 | -1.9 | | 29126106680 (38) | 0.01 | | | 29115876217 (26) | 0.03 | |
| | 29137023549 (46) | 0.05 | | | 29126122780 (14) | 0.04 | | P(17,9) | 29115953974.5 (6.0) | 0.22 | -4.0 |
| | 29137100843 (36) | 0.12 | | | 29126138213 (10) | 0.06 | | | 29116009525 (41) | 0.02 | |
| | 29137115133 (52) | 0.02 | | | 29126139343 (14) | 0.04 | | | 29116077351 (41) | 0.02 | |
| | 29137146870 (52) | 0.02 | | | 29126155121.4 (9.0) | 0.04 | | | 29116153143 (43) | 0.02 | |
| P(15,9) | 29137152722 (19) | 0.50 | -13.9 | | 29126173835 (12) | 0.04 | | P(17,10) | 29116154903.4 (5.6) | 0.34 | -10.2 |
| | 29137246904 (40) | 0.07 | | | 29126192390 (11) | 0.04 | | | 29116196315 (11) | 0.05 | |
| | 29137328958 (50) | 0.03 | | | 29126210044 (24) | 0.02 | | | 29116266702 (12) | 0.06 | |
| P(15,10) | 29137354496 (19) | 0.38 | 0.1 | P(16,7) | 29126216153.8 (5.4) | 0.59 | 0.9 | | 29116316174.7 (5.6) | 0.40 | |
| | 29137385638 (47) | 0.04 | | | 29126224518 (13) | 0.04 | | P(17,11) | 29116377183.0 (5.6) | 0.35 | 1.0 |
| | 29137387359 (54) | 0.01 | | | 29126228834 (16) | 0.03 | | | 29116420819.7 (6.6) | 0.21 | |
| | 29137427745 (48)* | 0.02 | | | 29126275613 (17) | 0.02 | | | 29116432668 (31) | 0.03 | |
| | 29137461256 (50)* | 0.02 | | | 29126352184 (23) | 0.02 | | | 29116457003 (19) | 0.04 | |
| | 29137526052 (50)* | 0.02 | | P(16,8) | 29126374897.0 (5.4) | 0.52 | -1.2 | | 29116494024 (41) | 0.02 | |
| | 29137562842 (51)* | 0.02 | | | 29126388114 (13) | 0.03 | | | 29116505568 (12) | 0.08 | |
| P(15,11) | 29137577673 (20)* | 0.32 | -1.4 | | 29126404753 (19) | 0.02 | | | 29116514259 (29) | 0.03 | |
| | 29137588489 (20)* | 0.21 | | | 29126406240.5 (7.8) | 0.06 | | | 29116536993 (18) | 0.05 | |
| | 29137614205 (43)* | 0.03 | | | 29126420721 (36) | 0.02 | | | 29116598710.2 (7.3) | 0.08 | |
| | 29137679345 (44)* | 0.03 | | | 29126448882 (43) | 0.01 | | P(17,12) | 29116620844.5 (5.5) | 0.51 | 6.4 |
| | 29137738945 (50)* | 0.02 | | | 29126451011 (28) | 0.02 | | | 29116647253 (12) | 0.04 | |
| | 29137739915 (37)* | 0.06 | | | 29126455713 (29) | 0.02 | | | 29116660342 (15) | 0.03 | |
| | 29137778805 (51)* | 0.02 | | | 29126466151 (42) | 0.01 | | | 29116701056.6 (6.4) | 0.12 | |
| P(15,12) | 29137822334 (25) | 0.20 | 5.7 | | 29126474707 (38) | 0.02 | | | 29116728624.2 (8.8) | 0.06 | |
| | 29137833166 (47) | 0.04 | | | 29126481373 (44) | 0.01 | | | 29116777697.4 (7.1) | 0.09 | |
| | 29137896745 (38) | 0.09 | | | 29126501107 (25) | 0.01 | | | 29116778298.9 (7.9) | 0.07 | |
| | 29137906138 (54) | 0.01 | | | 29126521861.4 (8.7) | 0.05 | | | 29116808069.3 (7.6) | 0.08 | |
| | 29137928937 (42) | 0.06 | | | 29126548046.3 (5.5) | 0.34 | | | 29116810371.7 (7.3) | 0.08 | |
| | 29137932072 (45) | 0.05 | | P(16,9) | 29126554923.5 (5.4) | 0.55 | -4.3 | | 29116813200.2 (7.1) | 0.09 | |
| | 29137938419 (43) | 0.06 | | | 29126591644 (37) | 0.01 | | | 29116829030.4 (8.4) | 0.06 | |
| | 29137945399 (53) | 0.01 | | | 29126602743 (14) | 0.03 | | | 29116833086 (23) | 0.02 | |

| | | | | | | | | | | |
|---|---|---|---|---|---|---|---|---|---|---|
| | 29137964056 (44) | 0.05 | | | 29126631269.0 (6.7) | 0.11 | | | 29116838609.9 (8.9) | 0.06 | |
| | 29137966867 (53) | 0.02 | | | 29126637688 (11) | 0.07 | | | 29116841389 (26) | 0.02 | |
| | 29137968611 (54) | 0.01 | | | 29126638842 (27) | 0.03 | | | 29116850637 (14) | 0.03 | |
| | 29137975251 (46) | 0.05 | | | 29126666124.2 (6.5) | 0.12 | | | 29116865389.9 (8.3) | 0.07 | |
| | 29137982460 (52) | 0.02 | | | 29126697167.5 (5.7) | 0.25 | | P(17,13) | 29116885949.8 (5.5) | 0.37 | 7.6 |
| | 29138004207 (48) | 0.03 | | | 29126707010 (36) | 0.01 | | | 29116902572 (11) | 0.04 | |
| | 29138017137 (53) | 0.02 | | | 29126733434.6 (6.5) | 0.10 | | | 29116904137.1 (5.4) | 0.48 | |
| | 29138035113 (44) | 0.05 | | | 29126746781.5 (6.8) | 0.09 | | | 29116933641 (11) | 0.06 | |
| | 29138037723 (47) | 0.04 | | P(16,10) | 29126756284.0 (5.4) | 0.57 | -1.6 | | 29116939164.1 (8.4) | 0.08 | |
| | 29138045534 (52) | 0.02 | | | 29126784865 (11) | 0.07 | | | 29116947125.4 (8.6) | 0.08 | |
| | 29138055674 (46) | 0.04 | | | 29126811465 (16) | 0.05 | | | 29116970167 (38) | 0.01 | |
| | 29138058891 (48) | 0.04 | | | 29126818678 (13) | 0.06 | | | 29116971832 (17) | 0.03 | |
| P(15,13) | 29138088525 (32) | 0.13 | 7.5 | | 29126843646.7 (7.8) | 0.12 | | | 29117004289.1 (7.8) | 0.10 | |
| | 29138104257 (53) | 0.02 | | | 29126915173 (15) | 0.05 | | | 29117020828 (21) | 0.03 | |
| | 29138113967 (51) | 0.03 | | | 29126926742 (32) | 0.02 | | | 29117031549.8 (8.6) | 0.08 | |
| | 29138123530 (47) | 0.04 | | P(16,11) | 29126979023.2 (5.5) | 0.48 | 1.9 | | 29117051832 (12) | 0.05 | |
| | 29138156588 (45) | 0.05 | | | 29127009834 (19) | 0.02 | | | 29117058467 (12) | 0.06 | |
| | | | | | 29127069846 (11) | 0.05 | | | 29117081562.2 (8.2) | 0.09 | |
| | | | | | 29127070792 (29) | 0.02 | | | 29117108210.7 (6.0) | 0.20 | |
| | | | | | 29127106810 (28) | 0.02 | | | 29117138755.8 (8.5) | 0.08 | |
| | | | | | 29127111934.5 (7.3) | 0.09 | | P(17,14) | 29117172571.1 (5.6) | 0.25 | 12.0 |
| | | | | | 29127124102.5 (7.8) | 0.08 | | | 29117178663 (12) | 0.04 | |
| | | | | | 29127141891 (30) | 0.01 | | | 29117241868.2 (8.5) | 0.08 | |
| | | | | | 29127165510 (36) | 0.01 | | | 29117258354 (12) | 0.05 | |
| | | | | | 29127200894.7 (6.7) | 0.07 | | | 29117259193 (11) | 0.06 | |
| | | | | | 29127218110.0 (7.6) | 0.05 | | | 29117327159 (12) | 0.05 | |
| | | | | P(16,12) | 29127223191.4 (5.5) | 0.39 | 1.9 | | 29117332600.7 (9.8) | 0.06 | |
| | | | | | 29127247897 (38) | 0.01 | | | 29117354610 (22) | 0.02 | |
| | | | | | 29127288257.2 (8.6) | 0.06 | | | 29117363851 (29) | 0.02 | |
| | | | | | 29127317510.6 (6.9) | 0.09 | | | 29117382173.5 (9.8) | 0.06 | |
| | | | | | 29127356251.7 (7.8) | 0.07 | | | 29117420983.8 (8.9) | 0.07 | |

| | | | | | | | | | | |
|---|---|---|---|---|---|---|---|---|---|---|
| | | | | | 29127358798.9 (7.3) | 0.08 | | | 29117457966 (18) | 0.03 | |
| | | | | | 29127369952.1 (5.7) | 0.21 | | | 29117478645 (11) | 0.04 | |
| | | | | | 29127389083.0 (7.8) | 0.07 | | | 29117479467 (13) | 0.03 | |
| | | | | | 29127392392.3 (8.3) | 0.06 | | P(17,15) | 29117480752.8 (5.9) | 0.16 | -6.2 |
| | | | | | 29127414618 (19) | 0.02 | | | 29117493690.8 (5.4) | 0.49 | |
| | | | | | 29127422978 (11) | 0.04 | | | 29117496998 (15) | 0.03 | |
| | | | | | 29127426696.6 (7.8) | 0.07 | | | 29117520977 (17) | 0.03 | |
| | | | | | 29127433300 (12) | 0.04 | | | 29117528487 (15) | 0.03 | |
| | | | | | 29127438729.9 (7.1) | 0.08 | | | 29117573763 (20) | 0.02 | |
| | | | | | 29127452351.9 (6.9) | 0.06 | | | 29117585680 (16) | 0.03 | |
| | | | | P(16,13) | 29127488856.3 (5.5) | 0.30 | 6.3 | | 29117619458.1 (7.7) | 0.07 | |
| | | | | | 29127502345 (24) | 0.02 | | | 29117638739.0 (7.3) | 0.08 | |
| | | | | | 29127525883 (26) | 0.01 | | | 29117650756 (11) | 0.04 | |
| | | | | | 29127530549 (20) | 0.02 | | | 29117725380.7 (7.6) | 0.07 | |
| | | | | | 29127532363 (17) | 0.02 | | | 29117732158.5 (7.4) | 0.08 | |
| | | | | | 29127536217.5 (7.4) | 0.07 | | | 29117737167 (16) | 0.03 | |
| | | | | | 29127550361.3 (7.1) | 0.07 | | | 29117785383.4 (5.6) | 0.24 | |
| | | | | | 29127553401 (30) | 0.01 | | | 29117787580.6 (8.7) | 0.06 | |
| | | | | | 29127564809.5 (7.6) | 0.06 | | | 29117799534.5 (7.8) | 0.07 | |
| | | | | | 29127588998 (22) | 0.02 | | | 29117806208.0 (7.0) | 0.09 | |
| | | | | | 29127617132.3 (6.8) | 0.08 | | P(17,16) | 29117810611.1 (7.5) | 0.07 | -6.5 |
| | | | | | 29127642668.2 (8.9) | 0.05 | | | | | |
| | | | | | 29127664040.1 (7.5) | 0.07 | | | | | |
| | | | | | 29127691012.3 (6.8) | 0.08 | | | | | |
| | | | | | 29127716840.2 (9.6) | 0.03 | | | | | |
| | | | | | 29127739184.0 (7.3) | 0.05 | | | | | |
| | | | | | 29127746602 (13) | 0.02 | | | | | |
| | | | | | 29127771022.7 (9.7) | 0.03 | | | | | |
| | | | | P(16,14) | 29127776070.9 (5.7) | 0.19 | 2.7 | | | | |
| | | | | | 29127781666 (12) | 0.04 | | | | | |
| | | | | | 29127794757.1 (9.5) | 0.04 | | | | | |

| | | | | | | | | | | |
|---|---|---|---|---|---|---|---|---|---|---|
| | | | | | 29127801878 (22) | 0.02 | | | | |
| | | | | | 29127812375.3 (9.2) | 0.05 | | | | |
| | | | | | 29127845221 (18) | 0.02 | | | | |
| | | | | | 29127850044.6 (7.3) | 0.07 | | | | |
| | | | | | 29127877141 (21) | 0.02 | | | | |
| | | | | | 29127888637 (23) | 0.02 | | | | |
| | | | | | 29127902398 (17) | 0.02 | | | | |
| | | | | | 29127914036 (20) | 0.02 | | | | |
| | | | | | 29127920849 (12) | 0.04 | | | | |
| | | | | | 29127921460 (23) | 0.02 | | | | |
| | | | | | 29127947585 (12) | 0.02 | | | | |
| | | | | | 29127974879.2 (7.5) | 0.06 | | | | |
| | | | | | 29127977960 (21) | 0.02 | | | | |
| | | | | | 29128025470.2 (8.3) | 0.05 | | | | |
| | | | | | 29128032086.6 (9.6) | 0.07 | | | | |
| | | | | | 29128032784.7 (5.8) | 0.26 | | | | |
| | | | | | 29128051453 (14) | 0.03 | | | | |
| | | | | $P(16,15)$ | 29128084906.9 (6.8) | 0.08 | -7.1 | | | |
| | | | | | 29128091865 (29) | 0.01 | | | | |
| | | | | | 29128097672.8 (7.5) | 0.06 | | | | |
| | | | | | 29128106460.0 (9.4) | 0.04 | | | | |
| | | | | | 29128149072.0 (9.2) | 0.05 | | | | |
| | | | | | 29128218299 (12) | 0.03 | | | | |
| | | | | | 29128226018.4 (7.9) | 0.06 | | | | |
| | | | | | 29128234827 (21) | 0.02 | | | | |
| | | | | | 29128236654.0 (6.5) | 0.09 | | | | |
| | | | | | 29128243444 (17.8) | 0.02 | | | | |